\documentclass{article}
\pdfoutput=1
\usepackage{amsmath,amssymb,graphicx}
\usepackage{framed}
\usepackage{microtype}
\usepackage{hyperref}
\usepackage{amsthm}
\usepackage{lscape}

\def\be{\begin{eqnarray}}
\def\ee{\end{eqnarray}}
\def\nn{\nonumber}

\def\p{\partial}

\def\Tr{{\rm Tr}\,}

\def\l[{\phantom.[}

\def\mq{Q}

\hoffset= - 1.0in         

\begin{document}

\title{\vspace{.1cm}{\Large {\bf $(q,t)$-KZ equation for
      Ding-Iohara-Miki algebra
}\vspace{.2cm}}
\author{
{\bf Hidetoshi Awata$^a$}\footnote{awata@math.nagoya-u.ac.jp},
\ {\bf Hiroaki Kanno$^{a,b}$}\footnote{kanno@math.nagoya-u.ac.jp},
\ {\bf Andrei Mironov$^{c,d,e}$}\footnote{mironov@lpi.ru; mironov@itep.ru},
\ {\bf Alexei Morozov$^{d}$}\thanks{morozov@itep.ru},\\
\ {\bf Andrey Morozov$^{d,e,f}$}\footnote{andrey.morozov@itep.ru},
\ {\bf Yusuke Ohkubo$^a$}\footnote{m12010t@math.nagoya-u.ac.jp}
\ \ and \ {\bf Yegor Zenkevich$^{d,g,h,i}$}\thanks{yegor.zenkevich@gmail.com}}
\date{ }
}

\maketitle

\vspace{-6.5cm}

\begin{center}
\hfill FIAN/TD-04/17\\
\hfill IITP/TH-08/17\\
\hfill ITEP/TH-014/17
\end{center}

\vspace{4.3cm}

\begin{center}
$^a$ {\small {\it Graduate School of Mathematics, Nagoya University,
Nagoya, 464-8602, Japan}}\\
$^b$ {\small {\it KMI, Nagoya University,
Nagoya, 464-8602, Japan}}\\
$^c$ {\small {\it Lebedev Physics Institute, Moscow 119991, Russia}}\\
$^d$ {\small {\it ITEP, Moscow 117218, Russia}}\\
$^e$ {\small {\it Institute for Information Transmission Problems, Moscow 127994, Russia}}\\
$^f$ {\small {\it Laboratory of Quantum Topology, Chelyabinsk State University, Chelyabinsk 454001, Russia }}\\
$^g$ {\small {\it Institute of Nuclear Research, Moscow 117312, Russia
  }}\\
$^h$ {\small {\it Dipartimento di Fisica, Universit\`a di Milano-Bicocca,
Piazza della Scienza 3, I-20126 Milano, Italy}}\\
$^i$ {\small {\it INFN, sezione di Milano-Bicocca,
I-20126 Milano, Italy
  }}
\end{center}

\vspace{.5cm}

\begin{abstract}
  We derive the generalization of the Knizhnik-Zamolodchikov equation
  (KZE) associated with the Ding-Iohara-Miki (DIM) algebra
  $U_{q,t}(\widehat{\widehat{\mathfrak{gl}}}_1)$. We demonstrate that
  certain refined topological string amplitudes satisfy these
  equations and find that the braiding transformations are performed
  by the $\mathcal{R}$-matrix of
  $U_{q,t}(\widehat{\widehat{\mathfrak{gl}}}_1)$. The resulting system
  is the uplifting of the $\widehat{\mathfrak{u}}_1$
  Wess-Zumino-Witten model. The solutions to the $(q,t)$-KZE are
  identified with the (spectral dual of) building blocks of the
  Nekrasov partition function for $5d$ linear quiver gauge
  theories. We also construct an elliptic version of the KZE and
  discuss its modular and monodromy properties, the latter being
  related to a dual version of KZE.
\end{abstract}

\bigskip

\bigskip

\section{Introduction}

The AGT relations \cite{AGT,AGT5d} connect two different worlds: that of
$2d$ conformal field theories \cite{CFT} and of instanton (SUSY ADHM)
moduli spaces \cite{ADHM,LMNS,Nakajima,Nekrasov}. The natural quantities in
the instanton world count embeddings of certain worldsheet manifold
$\Sigma$ into the moduli space $\mathcal{M}_{\mathbf{k}}$ of
instantons (e.g.\ for $\Sigma$ being Riemann surface, they are related to
Gromov-Witten invariants~\cite{GW}). In the simplest case of $\Sigma$
being a point, the generating functions of embeddings reduce to
(super)volumes and are expressed via Nekrasov functions \cite{Nekrasov,NO}. The quantities at
the CFT side are various conformal blocks. According to the AGT
relations~\cite{AGT}:
\begin{equation}
  {\rm
    Conformal\ blocks} = {\rm Nekrasov\ functions}
\end{equation}
More generally, one can consider embeddings of arbitrary worldsheets
which naturally carry more information about the moduli space (e.g.\
quantum multiplication).

Let us briefly remind the gauge theory interpretation of these quantities. The
path integral of the $4d$ $\mathcal{N}=2$ gauge theory localizes on
the instanton moduli space and can thus be reduced to the finite
dimensional LMNS integral over the ADHM data. This integral gives
the (regularized) volume of the moduli space and can itself be
localized by putting the theory in the $\Omega$-background
$\mathbb{R}^4_{\epsilon_1, \epsilon_2}$. This gives the Nekrasov
function, i.e.\ the cleverly regularized volume (more precisely,
equivariant cohomology) of the ADHM space:
\begin{equation}
  \label{eq:64}
  Z_{4d} = \sum_{\mathbf{k}} \zeta^{\mathbf{k}} \int_{\mathcal{M}_{\mathbf{k}}} 1 = Z_{\mathrm{Nekr}}
\end{equation}
Consider now the $5d$ $\mathcal{N}=1$ uplift of the $4d$ theory to
$\mathbb{R}^4_{q,t^{-1}} \times S^1$.  Configurations, on which the
path integral localizes, are now the \emph{maps} from $S^1$ to the
instanton moduli space. The integral over them is the index of
supersymmetric quantum mechanics with the target space
$\mathcal{M}_{\mathbf{k}}$, which can be again localized to a finite
dimensional LMNS integral, but the measure is generalized from
rational to trigonometric. The {\it index} can also can be understood
in terms of equivariant $K$-theory of the ADHM moduli space:
\begin{equation}
  \label{eq:64p}
  Z_{5d} = \sum_{\mathbf{k}} \zeta^{\mathbf{k}}
  \# \left(S^1 \to \mathcal{M}_{\mathbf{k}}\right) = \sum_{\mathbf{k}} \zeta^{\mathbf{k}}\, \mathrm{Ind}_{\mathcal{M}_{\mathbf{k}}} (q) = Z_{\mathrm{Nekr}}^{K\mathrm{-theor}}(q)
\end{equation}
We can now generalize the uplifting procedure to higher
dimensions. The partition function of the $6d$ theory on $T^2
\times \mathbb{R}_{\epsilon_1, \epsilon_2}^4$ naturally counts maps from $T^2$ to
the instanton moduli space. The partition function is given by the
{\it elliptic genus} of $\mathcal{M}_{\mathbf{k}}$:
\begin{equation}
  \label{eq:72}
  Z_{6d} = \sum_{\mathbf{k}} \zeta^{\mathbf{k}}
  \# \left(T^2 \to \mathcal{M}_{\mathbf{k}}\right) = \sum_{\mathbf{k}}
  \zeta^{\mathbf{k}} \, \mathrm{Ind}^{\mathrm{ell}}_{\mathcal{M}_{\mathbf{k}}} (q,p) = Z_{\mathrm{Nekr}}^{\mathrm{ell}}(q,p)
\end{equation}
One can consider further generalizations, with curved spaces and
embedded branes and more complicated subvarieties of different
dimensions, but the pattern is already clear: the higher-dimensional
partition functions (at least their low-energy limits)
can be thought of as generating functions of mappings from some target
to the ADHM moduli space.

Of course, the same is true if we extend our theories to liftings from
\emph{lower} dimensions: this reveals new structures hidden in the
naive $4d$ consideration.  One can consider the $5d$ theory on
$\mathbb{R}^4_{q,t^{-1}}\times S^1$ as a $3d$ theory with worldsheet $S^1\times
\mathbb{R}^2_q$ and target space being the moduli space
$\mathcal{V}_{\mathbf{k}}$ of $2d$ \emph{vortices.} This gives the
$3d$--$5d$ correspondence between the gauge
theories~\cite{Aganagic-triality}. The resulting integrals over
$\mathcal{V}_{\mathbf{k}}$ are manifestly equivalent to
Dotsenko-Fateev (DF) integrals in $q$-deformed CFT. Combining this
equivalence with spectral duality~\cite{spec-dual} of conformal blocks
we get the AGT duality between $5d$ gauge theories and $q$-deformed
CFT correlators.

\bigskip

We can summarize these basic relations between the CFT and ADHM objects in the following table:

\bigskip

\bigskip

\centerline{
{\footnotesize
\begin{tabular}{rcccc}
  &Dotsenko-Fateev&  &ADHM&\\
  &integrals&$\stackrel{ {\rm AGT} }{\longleftrightarrow}$&spaces $\mathcal{M}_{\mathbf{k}}$&\\
  &&&&\\
  &$\left<\prod \Psi\Psi^* \exp\left(\oint {\cal J}\right)  \right>$&
  & $J=\sum_{\bf k} \ \zeta^{\bf k}\cdot
  \#\Big(\Sigma \longrightarrow {\cal M}_{\bf k}\Big) $&\\
  &&&&\\
  &$W_{q,t}(\widehat{\mathfrak{gl}}_1) $ conformal block&&$\Sigma = S^1\times\mathbb{R}^2_{\mq}$ \ $\Longrightarrow$  &
  The setup of~\cite{Okounkov-Smirnov}
\\
&&&&\\
&$W_{\beta}(\widehat{\mathfrak{gl}}_1) $ conformal block &&$\Sigma =S^2$ \ $ \Longrightarrow$ & Genus-zero Gromov-Witten
invariants of $\mathcal{M}_{\mathbf{k}}$ \\
&&&&\\
single-line BNM \ $\subset$&Balanced network models&
&&\\
satisfies $\boxed{(q,t)\text{-KZE}}$&
$\cup$ &&&\\
&$W_{q,t}(\mathfrak{gl}_N)$ conformal blocks&&$\Sigma = S^1$ \ $ \Longrightarrow$ & $K$-theory of $\mathcal{M}_{\mathbf{k}}$ \\
&&&&\\
WZW$_{\mathfrak{u}_1}$\ $\subset$&Lioville/W models& &
$\Sigma = {\rm point}$\ $\Longrightarrow$ &
Volume (equivariant cohomology) of ${\cal M}_{\mathbf{k}}$ \\
satisfies KZE&conformal blocks&$\stackrel{ {\rm AGT} }{\longleftrightarrow}$
&&expressed through Nekrasov functions
\end{tabular}
}}

\bigskip

\bigskip

\noindent
The top line in the table can be thought of as the partition function
of the $7d$ gauge theory on $S^1 \times \mathbb{R}^6_{q,t^{-1},Q}$ (we
remind that the $\Omega$-background in the space of dimension $2n$ is
related to the action of $T^n$ on $\mathbb{C}^n$ and depends on $n$
deformation parameters).  Equivalently it corresponds to the $5d$
theory with matter content given by the quiver, which represents the
moduli for $2d$ vortices: one adjoint hypermultiplet plus several
fundamental ones. There is a third way of looking at this theory: it
can be understood as the $3d$ gauge sigma model on the ADHM moduli
space. It is this last option, that is reflected in the table. All the
three ($3d$--$5d$--$7d$) ways of looking at the theory are
complementary and useful in different situations. For example, from
the $7d$ perspective one can naturally see the symmetry between the
three equivariant parameters: two from the $\Omega$-background
$\mathbb{R}_{q, t^{-1}}^4$, in which the ADHM instantons live, and one
from $S^1 \times \mathbb{R}^2_Q$. There are several limits of this
construction which are also listed in the table.

The symmetry, which controls the properties of the models in this
pattern is the DIM algebra \cite{DIM,MMZpagoda,NagoyaITEP} and its further generalizations.
Already the simplest DIM algebra
$U_{q,t}(\widehat{\widehat{\mathfrak{gl}}}_1)$ involves a
$(q,t)$-deformation at least, which means that in the instanton (ADHM)
moduli space story, we make a lift from the level of $\Sigma=point$
and the \emph{volume} (equivariant cohomology) of the moduli space to,
at least, that of $\Sigma=S^1$ ($K$-theory of the ADHM space).
Further lifting to $\Sigma=$ Riemann surface should correspond to
further (elliptic) generalizations of DIM.  Pagoda algebra
$U_{q,t,p}(\widehat{\widehat{\widehat{\mathfrak{gl}}}}_1)$
from~\cite{MMZpagoda} is expected to be related to the most general
system of this kind.

Since the left low corner of this pattern is best understood, it is
most natural to look at the structures which are well known there and
try to extend them to other places.  From this perspective, the
network model generalizes the Dotsenko-Fateev (DF) realization
\cite{DF,MMSh} of $2d$ conformal blocks as correlators of the free
field vertex operators and screening charges, which commute with the
operator algebra (Virasoro, $W$ or Kac-Moody) and can be expressed as
integrals of screening currents.  {\it Balanced network model}
\cite{bn,NagoyaITEP} is built from a trivalent graph with lines of rational
slopes in such a way that all external ends of the graph are either
strictly horizontal or strictly vertical. Generally the lines are
associated with Fock representations of DIM algebra with the central
charges $(0,1)$ (vertical lines) or $(1,N)$ (horizontal lines). The
vertex operators, standing at the vertices, intertwine these
representations of DIM algebra~\cite{AFS,NagoyaITEP} and are expressed through
$(q,t)$-oscillators.  Moreover, exponentiated screening charges appear
automatically incorporated into the vertex operators, so that the
right number of screening charges is selected by conservation laws for
any concrete network.

\bigskip

{\bf The present paper} addresses the question what happens at the DIM
level, when the Liouville/$W_N$ model is substituted by the
Wess-Zumino-Witten (WZW) model \cite{WZNW,GMMOS}.  Usually, this means
that the operator algebra is extended from Virasoro or $W_N$ to a
Kac-Moody algebra $\widehat {\mathfrak{g}}$, e.g.\
$\widehat{\mathfrak{su}}_N$. Such an \emph{extension} of the symmetry
implies a \emph{reduction} of the set of screening charges:
 they should now commute not only with the stress-energy tensor and
 with $W$-operators, but also with all the currents of
 $\widehat{\mathfrak{g}}$. In the result, the set of WZW correlation
functions is a small subset of those in a generic CFT of the
Liouville/$W_N$ type.  As a manifestation of this, the correlators in
WZW model satisfy the universal first order Knizhnik-Zamolodchikov
equations~\cite{KZ,FR} (KZE), relating the shift in position of vertex
operators to a transformation in the group space, that is, to action
of the ${\cal R}$-matrix. KZE for DIM algebra can be schematically
written as follows:
\begin{equation}
  \label{eq:73}
 \text{``} \left( \frac{q}{t} \right)^{z_k \frac{\partial}{ \partial_{z_k}}} \langle V_1(z_1) \cdots V_N(z_N) \rangle
=  \prod_{i \neq k} {\mathcal R}_{i,k}  \langle V_1(z_1) \cdots
V_N(z_N) \rangle\, \text{''}
\end{equation}
where $V_i(z_i)$ are the  primary fields and $\mathcal{R}_{i,j}$ is
the $\mathcal{R}$-matrix acting on the $i$-th and $j$-th fields.

Let us recall that in general, the number of free fields in the
bosonization of the WZW model is ${\rm dim}(\mathfrak{su}_N)\sim N^2$
\cite{GMMOS}. The usual $W_N$-models are, in fact, {\it reductions} of
${\rm WZW}_{\mathfrak{su}_N}$.  For lowest Kac-Moody central charge
$k=1$, the number of the free fields in the bosonization reduces to
$(N-1)$. In such a description the WZW screening currents are
restricted to $\exp(\vec\alpha_i\cdot\vec\phi)$ with $\alpha_i$ being
the simple roots, while the Virasoro screenings are restricted by the
only condition $\vec\alpha^2=2$.

\begin{figure}[h]
  \centering
      \parbox{13cm}{\includegraphics[width=13cm]{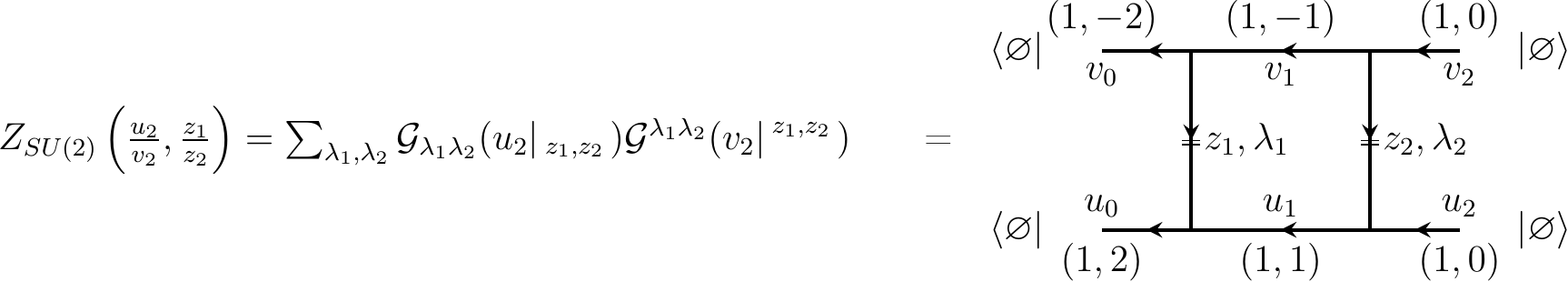}}
      \caption{The combination of two horizontal strings of
        intertwiners giving the pure $SU(2)$ Nekrasov function. The
        AGT dual of this function is the four-point Virasoro conformal
        block.}
  \label{Nekr}
\end{figure}

Another thing to keep in mind is that the simplest DIM algebra
considered in \cite{NagoyaITEP}, was
$U_{q,t}(\widehat{\widehat{\mathfrak{gl}}}_1)$ rather than
$\widehat{\widehat{\mathfrak{g}}}$ for a simple $\mathfrak{g}$ like
$\mathfrak{su}_N$.  Thus, it retains some Abelian traits, which
simplify the WZW theory considerably.  In the Kac-Moody case, the
``Abelian'' ${\rm WZW}_{\mathfrak{u}_1}$ model associated with the
(non-Abelian) algebra $\widehat{\mathfrak{u}}_1$ is nearly trivial:
only free field correlators without any screenings are picked up and
satisfy the corresponding KZE (see sec.~\ref{sec:kze-kac-moody} for
this archetypical example) though the KZE itself does not look much
simpler in this case. The lift of this ``Abelian'' model to the DIM
level is large enough to include all the $W_N$ models with any $N$ and
with all their generalized hypergeometric correlators. In this sense
the introduction of an additional loop is equivalent to considering
all $W_N$ in a unified way.  However, in the balanced network model
there is a remnant of $N$: the number of horizontal lines, see
Fig.~\ref{Nekr}.

The number of lines fixes the number of Virasoro/$W_N$-screenings,
which are actually associated with the vertical segments between the
horizontal lines.  Therefore, in the model with just a single
horizontal line there are no screenings at all, and thus it is the
proper lifting ($(q,t)$-deformation) of the WZW$_{\mathfrak{u}_1}$.
It is natural then to expect that it satisfies a $(q,t)$-deformed
version of KZE. In sec.~\ref{sec:extens-dim-wideh} we demonstrate that
this is indeed the case.

Moreover, one can derive the $(q,t)$-KZE for conformal blocks on an
elliptic curve to obtain a counterpart of the
Knizhnik-Zamolodchikov-Bernard equation (KZBE)
\cite{KZB}. Technically, this is done by considering the traces
(partition functions) instead of matrix elements (correlation
functions)~\cite{Etin} and by applying a counterpart of the
``averaging'' procedure \cite{RF}, which ultimately leads to the
emergence of the elliptic ${\cal R}$-matrix. We deal with the KZBE in
section~\ref{sec:elliptic-q-kz}, and derive, along with the KZBE, also
a monodromy equation, which turns out to be dual to the KZE (cf.\ the
quantum affine case in \cite{Etin, Sun}).

From above explanations, it is clear that this $(q,t)$-KZE is
\emph{not} a generalization of ``truly non-Abelian'' equations,
neither the original (classical) KZE, nor its quantum ($q$)
deformation from \cite{FR}: in order to derive their liftings, one
should consider $\widehat{\widehat{\mathfrak{g}}}$ with non-Abelian
$\mathfrak{g}$.  However,
$U_{q,t}(\widehat{\widehat{\mathfrak{gl}}}_1)$ is large enough to
possess a non-trivial ${\cal R}$-matrix \cite{Sm,NagoyaITEP3}, and the
$(q,t)$-KZE satisfied by the correlators in this case is also
non-trivial. A reminiscence of its Abelian origin is simplicity of the
${\cal R}$-matrix: in fact, though being infinite-dimensional, it is
diagonal and belongs to the set of trigonometric ${\cal R}$-matrices,
where spectral parameter can not be excluded without trivializing the
matrix itself.  As a manifestation of this, such ${\cal R}$-matrices
can \emph{not} be used in knot calculus: they satisfy the relation
\cite{NagoyaITEP3}
\begin{equation}
  {\cal R}^{-1}(z,w) ={\cal R}(w,z)
\end{equation}
which make the associated knot polynomials trivial.  Thus, for knot
theory purposes further lifting to $\widehat{\widehat{\mathfrak{g}}}$
is essential. However, the $(q,t)$-KZE (\ref{eq:30})--(\ref{eq:31}) by
itself is a nice non-trivial equation.

It deserves noting that the difference equations derived by A. Okounkov
and A. Smirnov~\cite{O,Okounkov-Smirnov,S} for the entries at the upper right corner
of the table above are quite different and involve very different
${\cal R}$-matrices. A much simpler and elementary equations
(\ref{eq:30})-(\ref{eq:31}) do not seem to appear in those papers.  It
is also an interesting open problem to clarify the relation of the
equations from~\cite{O,Okounkov-Smirnov,S} to the $(q,t)$-KZE from the viewpoint of
AGT relation.

For the detailed description of network models we refer the reader to
\cite{bn,MMZ,MMZpagoda,NagoyaITEP}, and for the associated
$RTT$-relations to \cite{NagoyaITEP2}.

\section{Warm-up: KZE for the Kac-Moody algebra $\widehat{\mathfrak{u}}_1$}
\label{sec:kze-kac-moody}

Let us explain the logic of constructing the KZE and its solutions in
the simplest example of the Kac-Moody algebra $\widehat
{\mathfrak{g}}$ \cite{KZ}. The whole story consists of three steps.
\begin{itemize}
\item{\underline{Step 1.}}  The derivation of KZE is based on the
  Sugawara construction, which expresses the stress tensor ${ T}(z)$
  through the currents ${ J}^a(z)$ of $\widehat{\mathfrak{g}}$:
  \begin{equation}
    \label{Sug}
    { T}(z) = \frac{1}{2} :{ J}^a{ J}^a(z):
  \end{equation}
  One can notice that the $L_{-1}$-generator acts as a derivative on
  the fields: $L_{-1}=\oint dz { T}(z)\to \partial/\partial z$. Hence,
  one can insert into the correlator the integral $\oint dz ...$ of
  equation (\ref{Sug}) and obtain some relation between correlators.
\item{\underline{Step 2.}}  For the correlator of primary fields of
  the current algebra, the insertion of the current is expressed
  through the correlator itself and described by the Ward identity
  \cite{KZ}
  \begin{equation}
    \left< { J}^a(z)V_i(z_i)\right>=\sum_i \frac{\rho_i(t^a)}{ z-z_i}\left< { J}^a(z)V_i(z_i)\right>
  \end{equation}
  where $\rho_i(t^a)$ are generators of $\mathfrak{g}$ acting on the
  $i$-th primary field $V_i$ which transforms in the representation
  $\rho_i$. Thus, for the correlators of this kind, (\ref{Sug}) gives
  a closed equation, which is called KZE.
\item{\underline{Step 3.}}  The KZE can be solved with the help of the
  free field formalism of \cite{GMMOS}, so that the primary fields are
  exponentials of free fields. However, it is a quite poor class of
  solutions. In order to have an ample class of solutions, one has to
  apply the Dotsenko-Fateev trick \cite{DF} and insert into the
  correlator a series of screening charges, the operators commuting
  with any generators (currents) ${ J}^a(z)$ of
  $\widehat{\mathfrak{g}}$, not only with the stress tensor. In order
  to construct them, one may take the screening currents, their
  integrals being screening charges. Then, the solutions to the KZE is
  given by Dotsenko-Fateev integrals made from ${\rm
    dim}(\mathfrak{g})$ free fields, the integrands being constructed
  from exponentials of these fields \cite{SchV}.
\end{itemize}
For the Kac-Moody algebra $\widehat
{\mathfrak{g}}=\widehat{\mathfrak{su}}_N$ with the central charge
$k=1$, the number of free fields can be reduced to ${\rm
  rank}(\mathfrak{g})=N-1$ and the screening currents are just
$e^{\vec\alpha_i\vec\phi}$ with $\vec\alpha_i$ being simple roots of
$\mathfrak{g}$, while {\it any} $\oint e^{\vec\alpha\vec\phi}$ with
$\vec\alpha^2=2$ commutes with ${ T}$, i.e. can be used as a Virasoro
screening charge.  Other (not only simple) roots can also be used, but
they do not produce new (independent) correlators.

In the simplest case of $\widehat
{\mathfrak{g}}=\widehat{\mathfrak{u}}_1$, there are no currents
commuting with the $\mathfrak{u}_1$-current, and all the correlators
in WZW$_{\mathfrak{u}_1}$ are just $\prod_{i<j} (z_i-z_j)^{a_ia_j}$,
while those of the single free field (Liouville model) can be multiple
integrals of such quantities, i.e.\ generalized hypergeometric
functions.

Indeed, the KZE in this case are
\be
\frac{\p F\{z|\alpha\}}{\p z_i} =
\sum_{j\neq i} \frac{\alpha_i\alpha_j}{z_i-z_j}\cdot F\{z|\alpha\}
\ee
where $F\{z|\alpha\}$ is a correlator of $V_{\alpha_i}(z_i)$
$\widehat{\mathfrak{u}}_1$-primaries. The most general solution to these equations has the form
\be
F\{z|\alpha\} \ \sim \ \prod_{i<j} (z_i-z_j)^{\alpha_i\alpha_j} \ \sim \left< e^{\alpha_i\phi(z_i)}\right>
\label{corrU1}
\ee
i.e.\ they do not admit any screening charges.

Note that the usual correlators in the Liouville theory are constructed using the Virasoro screening charge $e^{b\phi(x)}$, where $b$ parameterizes the central charge $c=1-6(b-1/b)^2$:
\be
F\{z|\alpha\} = \left< \prod_i e^{\alpha_i\phi(z_i)}\prod_\mu \oint e^{b\phi(x_\mu)}dx_\mu \right>
\ \sim \ \prod_{i<j} (z_i-z_j)^{\alpha_i\alpha_j}\cdot
\oint dx_\mu  \prod_{\mu>\nu} (x_\mu-x_\nu)^{b^2} \prod_{i,\mu}
(z_i-x_\mu)^{b\alpha_i}
\label{corrL1}
\ee
This correlator, however, does not satisfy the KZE because the variation of $z_i$ is induced by the action
of the stress tensor $T(z)$, which has a peculiar form of the square
of the $\widehat{\mathfrak{u}}_1$ current $J(z)=\partial\phi(z)$:
\be
\frac{\p F}{\p z_i} = \left<\oint_{z_i} T(y)dy
\prod_j e^{\alpha_j\phi(z_j)}\prod_\mu \oint e^{b\phi(x_\mu)}dx_\mu \right>
= -\sum_{k\neq i}  \left<\oint_{z_k} T(y)dy
\prod_j e^{\alpha_j\phi(z_j)}\prod_\mu \oint e^{b\phi(x_\mu)}dx_\mu \right>
=\nn \\
= -\sum_{k\neq i} \left<
\underbrace{\oint_{z_k} T(y)dy e^{\alpha_k\phi(z_k)} }
_{:\alpha_k\p\phi(z_k) e^{\alpha_k\phi(z_k)}:}
\prod_{j\neq k} e^{\alpha_j\phi(z_j)}\prod_\mu \oint e^{b\phi(x_\mu)}dx_\mu
\right>
= -\underbrace{\sum_{k\neq i}\sum_{j\neq k} \frac{\alpha_j\alpha_k}{z_j-z_k}}
_{\sum_{j\neq i}  \frac{\alpha_i\alpha_j}{z_i-z_j}}
\cdot F
+ \parbox{3cm}{contributions from pairings with screenings}
\label{KZU1}
\ee
In fact, the last contribution can not be neglected in the $\widehat{\mathfrak{u}}_1$ case:
already the hypergeometric $F$ (for one screening) satisfies a second order
differential equation rather than the first order KZE.
Thus, the WZW model for $\widehat {\mathfrak{g}}=\widehat{\mathfrak{u}}_1$ has only correlators
of the form (\ref{corrU1}) and not the generic Liouville theory
correlators (\ref{corrL1}). At the same time, the Sugawara equation (\ref{Sug}) being inserted into(\ref{corrL1}) gives rise to some relations between {\it different} correlators in the generic Liouville theory.

What happens for {\it simple} algebras $\mathfrak{g}$ is that there is
a restricted, still non-empty set of screenings, which commute with
the currents, not only with the stress tensor \cite{GMMOS}, and the
analogue of the last contribution in (\ref{KZU1}) is absent for them.
This selects correlators of the WZW model, which is a subset of the
correlators of the associated $W(\mathfrak{g})$ (Toda) model.

\bigskip

One can similarly derive the KZE on torus. In the $\widehat{\mathfrak{u}}_1$ case, the equation has the form
\be
\frac{\p F\{z|\alpha\}}{\p z_i} =
\sum_{j\neq i} \alpha_i\alpha_jr(z_i-z_j)\cdot F\{z|\alpha\}
\ee
where $r(z)$ is the logarithmic derivative of the $\theta$-function: $r(z)=\left[\log\theta (z)\right]'$ and the solution is rather trivial:
\be
F\{z|\alpha\} \ \sim \ \prod_{i<j} \theta(z_i-z_j)^{\alpha_i\alpha_j}
\ee
The extension of this equation to arbitrary $\widehat{\mathfrak{g}}$ and the Riemann surface of arbitrary genus is rather involved and sometimes called KZBE \cite{KZB}.

\bigskip

The ($q$-)Knizhnik-Zamolodchikov equation extends these results from
$\widehat{\mathfrak{u}}_1$ to arbitrary Kac-Moody algebra, classical
and quantum. The present paper initiates the program of extending it
further to the DIM algebra. To this end, we need to translate the
above properties into a pure algebraic language, with vertex operators
interpreted as intertwiners of Fock representations and construct a
counterpart of the Sugawara relation.  However, the stress tensor and
operator expansions are, in fact, not necessarily needed: one needs
just the shift operator expressed through currents and appropriate
commutation relations. Moreover, the substitute of the Sugawara
construction for the shift operator is just a decomposition \be
\label{DIMSug} \Psi(pz) = {\cal T}_+\Psi(z){\cal T}_-
\ee
where the
subscripts $\pm$ denote the parts with only positive or negative
powers of the spectral parameter. In practice, they are realized by
taking the DIM ${\cal T}$-operator at special values of spectral
parameters. Let us stress here that ${\cal T}$ is the resolved
conifold, i.e.\ a bilinear combination of two topological vertices
(intertwiners) and {\bf is not} the energy-momentum tensor.

Eq.~(\ref{DIMSug}) is the DIM counterpart of $L_{-1} =
J^a_+J^a_-$. Hence, with this equation we only have step 1 done. In
order to make step 2, we find the commutation relations of ${\cal T}$
and $\Psi$, and, as step 3, one obtains the algebraic solutions
similarly to the $\widehat{\mathfrak{u}}_1$ case. Similarly to the
$\widehat{\mathfrak{u}}_1$ case, there are no nontrivial screening operators for
$U_{q,t}(\widehat{\widehat{\mathfrak{gl}}}_1)$.

\section{Extension to DIM algebra $U_{q,t}(\widehat{\widehat{\mathfrak{gl}}}_1)$ }
\label{sec:extens-dim-wideh}

There are two (different) directions to generalize the story of
$\mathfrak{u}_1$ WZW model, or Abelian KZ equation. One is to the
quantum affine algebra $U_q(\widehat{\mathfrak{sl}}_N)$ and the other
is to DIM algebra $U_{q,t}(\widehat{\widehat{\mathfrak{gl}}}_1)$.  In
the first case non-Abelian nature of $\mathfrak{sl}_N$ allows the
screening operators which make the correlation functions of
hypergeometric type. On the other hand in the second case we keep
Abelian nature of the algebra. The structure of correlation functions
is of free field type typically obtained by use of the Wick
contraction, but the building block (the ``propagator'') becomes
rather non-trivial and interesting.  Still there are several common
features.  We have a free field realization of the Drinfeld currents
in both cases. The quantum deformation of KZ equation for the
correlation functions of the intertwining operators (vertex operators)
takes the similar form which involves a product of $\mathcal{R}$
matrices. Hence, we can expect these two cases (addition of the second
loop and non-Abelian generalization) are unified by considering the
quantum toroidal algebra $U_{q,t}(\widehat{\widehat{\mathfrak{g}}})$
for non-Abelian algebra $\mathfrak{g}$.

In this section, we derive and solve the KZE for the DIM algebra
$U_{q,t}(\widehat{\widehat{\mathfrak{gl}}}_1)$. We proceed in
steps. First, in sec.~\ref{sec:difference-operator} we rewrite the
shift operator acting on the vertical spectral parameter of the
intertwiner $\Psi$ in terms of the $\mathcal{T}$-operators,
schematically
\begin{gather}
  \label{eq:25}
  ``\Psi \left( \frac{q}{t} z \right) = \mathcal{T}(z) \Psi(z)
  \mathcal{T}^{-1}(z)"\\
 \parbox{12cm}{\includegraphics[width=12cm]{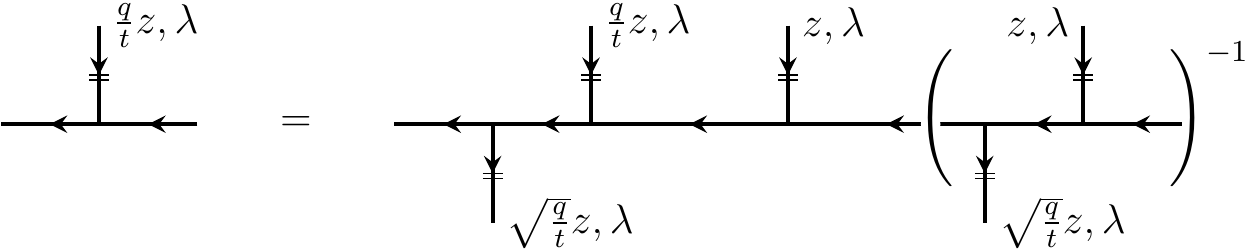}}
\end{gather}
where the $\mathcal{T}(z)$ operators are evaluated for certain
concrete values of the spectral parameters depicted on the
figure. This can be viewed as the DIM version of the Sugawara
construction which relates the conformal and algebraic properties of
the intertwiners. Then in sec.~\ref{sec:commutation-vertex-t} we
derive the commutation relations between the $\mathcal{T}$-operators
and the intertwiners, which look like:
\begin{equation}
  \label{eq:26}
  ``\mathcal{T} \Psi =  \mathcal{R} \Psi \mathcal{T} "
\end{equation}
with $\mathcal{R}$ representing the $\mathcal{R}$-matrix. In
sec.~\ref{sec:q-kz-equation}, we combine the obtained relations and
write down the $(q,t)$-KZ equation for the vacuum matrix element of the
product of several intertwiners of the form
\begin{multline}
  \label{eq:27}
  ``\langle \varnothing | \Psi \cdots \Psi \left( \frac{q}{t} z \right)
  \cdots \Psi | \varnothing \rangle = \langle \varnothing | \Psi
  \cdots \mathcal{T} \Psi (z) \mathcal{T}^{-1}
  \cdots \Psi | \varnothing \rangle = \left(\prod \mathcal{R} \right) \langle \varnothing |\mathcal{T} \Psi
  \cdots  \Psi (z)
  \cdots \Psi \mathcal{T}^{-1} | \varnothing \rangle =\\
  =\left(\prod \mathcal{R} \right) \langle \varnothing | \Psi
  \cdots  \Psi (z)
  \cdots \Psi  | \varnothing \rangle "
\end{multline}
where in the last line we have used the fact that the $\mathcal{T}$
operators which appear for the special values of the spectral parameters
act trivially on the vacuum vectors. Finally, in
sec.~\ref{sec:solution-q-kz}, we give explicit solutions to the DIM
$(q,t)$-KZ equation. This solution is similar to the solution of
$\widehat{\mathfrak{u}}_1$ KZE. It is a product of two-point
correlators (again, schematically)
\begin{equation}
  \label{eq:28}
  ``\langle \varnothing | \Psi(z_1)
  \cdots \Psi(z_n)  | \varnothing \rangle = \prod_{i<j}   \langle
  \varnothing | \Psi(z_i)  \Psi(z_j) | \varnothing \rangle "
\end{equation}
or, more generally,
\begin{equation}
  \label{eq:29}
  ``\langle \varnothing |\Psi^{*}(w_1)
  \cdots \Psi^{*}(w_m) \Psi(z_1)
  \cdots \Psi(z_n)  | \varnothing \rangle = \prod_{k<l}   \langle
  \varnothing | \Psi^{*}(w_k)  \Psi^{*}(w_l) | \varnothing \rangle \prod_{i<j}   \langle
  \varnothing | \Psi(z_i)  \Psi(z_j) | \varnothing \rangle \prod_{k,i}  \langle
  \varnothing | \Psi^{*}(w_k)  \Psi(z_i) | \varnothing \rangle "
\end{equation}

\subsection{Difference operator}
\label{sec:difference-operator}

Let us first derive the operator which produces the shift
in the vertical spectral parameters $(z,w)$ of the intertwiners
$\Psi^\lambda$ and $\Psi^{*}_\mu$, where $\lambda$ and
$\mu$ are partitions attached to the vertical line.
The intertwiners also depend on the horizontal spectral
parameter $u$ and on the slope parameter $N$.
Let us recall that the explicit expression for them is given in \cite{AFS,NagoyaITEP},
\begin{multline}
  \label{eq:23}
  \Psi^{\lambda}(N,u|z)= \parbox{3cm}{\includegraphics[width=3cm]{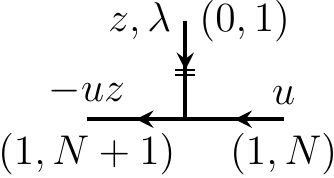}} = (-z)^{-N|\lambda|} u^{|\lambda|}
  f_{\lambda}^{-N-1} \frac{q^{n(\lambda^{\mathrm{T}})}}{C_{\lambda}}
  \exp \left[ \sum_{n \geq 1} \frac{z^n}{n} \left( - \frac{1}{1-q^n} +
      (1-t^{-n}) \mathrm{Ch}_{\lambda}(q^n, t^n) \right) a_{-n}
  \right]\times\\
  \times \exp \left[ \sum_{n \geq 1} \frac{z^{-n}}{n} \left( - \frac{q^n}{1-q^n} -
      (1-t^n) \mathrm{Ch}_{\lambda}(q^{-n}, t^{-n}) \right) a_n
  \right]
\end{multline}
and
\begin{multline}
\label{eq:56}
\Psi^{*}_{\mu}(N,u|w)= \parbox{3cm}{\includegraphics[width=3cm]{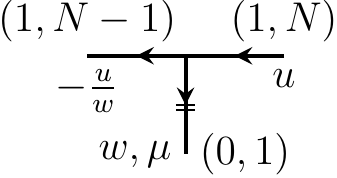}}
= (-w)^{N|\mu|} u^{-|\mu|} f_{\mu}^{N-1}
\frac{q^{n(\mu^{\mathrm{T}})+|\mu|}}{C_{\mu}} \times\\
\times\exp \left[ \sum_{n \geq 1}
   \frac{z^n}{n} \left( \frac{t}{q} \right)^{\frac{n}{2}} \left( \frac{1}{1-q^n} - (1-t^{-n})
    \mathrm{Ch}_{\mu}(q^n, t^n) \right) a_{-n}
\right]
\exp \left[  \sum_{n \geq 1} \frac{z^{-n}}{n} \left(
    \frac{t}{q} \right)^{\frac{n}{2}} \left(
    \frac{q^n}{1-q^n} + (1-t^n) \mathrm{Ch}_{\mu}(q^{-n}, t^{-n})
  \right) a_n \right]
\end{multline}
Here and henceforth, we use the following notations:
\begin{gather}
  \label{eq:24}
  \vert \lambda \vert = \sum_{i} \lambda_i\\
  f_{\lambda} = (-1)^{|\lambda|} q^{n(\lambda^{\mathrm{T}}) +
    \frac{|\lambda|}{2}} t^{-n(\lambda) - \frac{|\lambda|}{2}},\qquad
  n(\lambda) = \sum_{(i,j) \in \lambda} (i-1) =
  \frac{||\lambda^{\mathrm{T}}||^2 - |\lambda|}{2},\\
  C_{\lambda} = \prod_{(i,j)\in \lambda} \left( 1 - q^{\lambda_i -j}
    t^{\lambda^{\mathrm{T}}_j - i + 1} \right), \qquad
  \mathrm{Ch}_{\lambda}(q,t) = \sum_{(i,j) \in \lambda} q^{j-1}
  t^{1-i}.
\end{gather}
where we identify a partition $\lambda = (\lambda_1 \geq \lambda_2
\geq \cdots \lambda_i \geq \cdots)$ with a Young diagram
and $(i,j) \in \lambda$ means a box in the diagram.
The Heisenberg algebra is
\be
[a_n,a_m] = n{1-q^{|n|}\over 1-t^{|n|}}\delta_{m+n,0}
\ee
We will omit the slope labels $(0,1)$ on the vertical lines and label
only the horizontal ones.

For general complex number $p$, one can write two difference equations
for the intertwiner: one in the horizontal and one in the vertical spectral
parameters. The former is trivial:
\begin{equation}
  \label{eq:11}
  \boxed{\Psi^{\lambda}(N,pu|z) = p^{|\lambda|} \Psi^{\lambda}(N,u|z)}
\end{equation}
It can be understood as the action of the vertical grading operator
$p^{d_{\perp}}$ on the Fock representations being intertwined \cite{AFS}. This
operator shifts the spectral parameters of the two horizontal legs of
the intertwiner and counts the level of the basis vectors in the
vertical leg, hence Eq.~\eqref{eq:11}.

The second equation is more involved and will be the one used in the
derivation of the $(q,t)$-KZ equation. For any $p$, one has
\begin{multline}
  \label{eq:1}
  \Psi^{\lambda}(N,u|pz) = p^{-N|\lambda|} \exp \left[ \sum_{n \geq 1} \frac{z^n}{n}
    (p^n-1) \left( - \frac{1}{1-q^n} + (1-t^{-n})
      \mathrm{Ch}_{\lambda} (q^n,t^n) \right) a_{-n} \right]
  \Psi^{\lambda}(N,u|z)\times\\
  \times\exp \left[ \sum_{n \geq 1} \frac{z^{-n}}{n} (p^{-n}-1) \left(
      - \frac{q^n}{1-q^n} - (1-t^n) \mathrm{Ch}_{\lambda}
      (q^{-n},t^{-n}) \right) a_n\right]
\end{multline}
It will be crucial for us that, at the special values of $p = \left(
  \frac{q}{t} \right)^{\pm 1}$, the two exponentials appearing in the
r.h.s.\ can be recast into the $\mathcal{T}$-operators, bilinear combinations of the intertwiners which geometrically
correspond to resolved conifolds (at the level of refined amplitudes) and satisfy the $\mathcal{RTT}$ relations with
the DIM algebra $\mathcal{R}$-matrix, see s.\ref{sec:commutation-vertex-t} below and \cite{NagoyaITEP2,NagoyaITEP3}. For
definiteness, consider $p = \frac{q}{t}$.

The general expression that defines the $\mathcal{T}$-operator is
\begin{multline}\label{30}
  \mathcal{T}^\lambda_\mu(N,u|z,w) = \parbox{3.5cm}{\includegraphics[width=3.5cm]{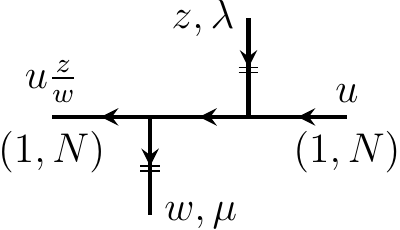}} = \Psi^{*}_{\mu}(N+1,-zu|w)
  \Psi^{\lambda}(N,u|z)=\\
  = (-z)^{-N|\lambda|} (-w)^{N|\mu|} u^{|\lambda|} \left( \frac{q w}{u
      z} \right)^{|\mu|} \frac{f_{\mu}^N q^{n(\lambda^{\mathrm{T}}) +
      n(\mu^{\mathrm{T}})} G_{\lambda \mu}^{(q,t)} \left(
      \sqrt{\frac{q}{t}} \frac{z}{w} \right)}{C_{\lambda} C_{\mu} f_{\lambda}^{N+1}} \exp
  \left[ \sum_{n \geq 1} \frac{1}{n} \left( \sqrt{\frac{q}{t}}
      \frac{z}{w} \right)^n \frac{1}{(1-q^n)(1-t^{-n})}  \right]
  \times\\
  \times \exp
  \left[ \sum_{n \geq 1} \frac{1}{n} \left\{ z^n \left( -
      \frac{1}{1-q^n} + (1-t^{-n})
      \mathrm{Ch}_{\lambda} (q^{-n},t^{-n}) \right) -  \left(
      \sqrt{\frac{t}{q}} w \right)^n \left( -
      \frac{1}{1-q^n} + (1-t^{-n})
      \mathrm{Ch}_{\mu} (q^n,t^n) \right)\right\} a_{-n} \right] \times\\
\times \exp
  \left[ \sum_{n \geq 1} \frac{1}{n} \left\{ z^{-n} \left( -
      \frac{q^n}{1-q^n} - (1-t^n)
      \mathrm{Ch}_{\lambda} (q^n,t^n) \right) -  \left(
      \sqrt{\frac{q}{t}} w \right)^{-n} \left( -
      \frac{q^n}{1-q^n} - (1-t^n)
      \mathrm{Ch}_{\mu} (q^{-n},t^{-n}) \right)\right\} a_n \right]
\end{multline}
where
\be
 G_{\alpha \beta}^{(q,t)}(x) = \prod_{(i,j)\in \alpha} \left( 1 - x
    q^{\alpha_i - j} t^{\beta^{\mathrm{T}}_j-i+1} \right)
  \prod_{(i,j)\in \beta} \left( 1 - x q^{-\beta_i + j-1} t^{-
      \alpha^{\mathrm{T}}_j+i} \right)
\ee
is the standard Nekrasov factor.

For $\frac{w}{z} = \left( \frac{q}{t} \right)^{\pm \frac{1}{2}}$ and
$\lambda = \mu$, (\ref{30}) simplifies as follows:
\begin{gather}
  \mathcal{T}^{\lambda}_{\lambda} \left(N, u \Big| z ,
    \sqrt{\frac{q}{t}} z \right) = \left( \frac{q}{t} \right)^{\frac{1}{2} (N+1) \vert\lambda\vert}
\frac{q^{|| \lambda ||^2} G_{\lambda\lambda}^{(q,t)}(1)}{C_\lambda^2
  f_\lambda}
\exp
  \left[ \sum_{n \geq 1} \frac{1}{n} \frac{1}{(1-q^n)(1-t^{-n})}
  \right]\times \notag\\
  \times \exp \left[ \sum_{n \geq 1} \frac{z^{-n}}{n} \left( 1 -
      \left( \frac{t}{q} \right)^n \right) \left( - \frac{q^n}{1-q^n}
      - (1-t^n)
      \mathrm{Ch}_{\lambda} (q^{-n},t^{-n}) \right) a_n
  \right],\label{eq:2}
  \\
  \notag\\
  \mathcal{T}^{\lambda}_{\lambda} \left(N, u \Big| z ,
    \sqrt{\frac{t}{q}} z \right) =
\left( \frac{q}{t} \right)^{- \frac{1}{2} (N+1) \vert\lambda\vert}
\frac{q^{|| \lambda ||^2} G_{\lambda\lambda}^{(q,t)}(q/t)}{C_\lambda^2 f_\lambda}
  \exp \left[ \sum_{n \geq 1} \frac{1}{n} \frac{\left( \frac{q}{t}
      \right)^n}{(1-q^n)(1-t^{-n})} \right]\times \notag\\
  \times \exp \left[ \sum_{n \geq 1} \frac{z^n}{n} \left( 1 - \left(
        \frac{t}{q} \right)^n \right) \left( - \frac{1}{1-q^n} +
      (1-t^{-n}) \mathrm{Ch}_{\lambda} (q^n,t^n) \right) a_{-n}
  \right]\label{eq:3}
\end{gather}
Notice that the $\mathcal{T}$-operators for the special values of
parameters in Eq.~\eqref{eq:2} (resp.~\eqref{eq:3}) contain only
bosonic annihilation (resp.\ creation) operators. Therefore, they act
trivially on the vacuum vectors:
\begin{gather}
  \label{eq:66}
  \parbox{14cm}{\includegraphics[width=14cm]{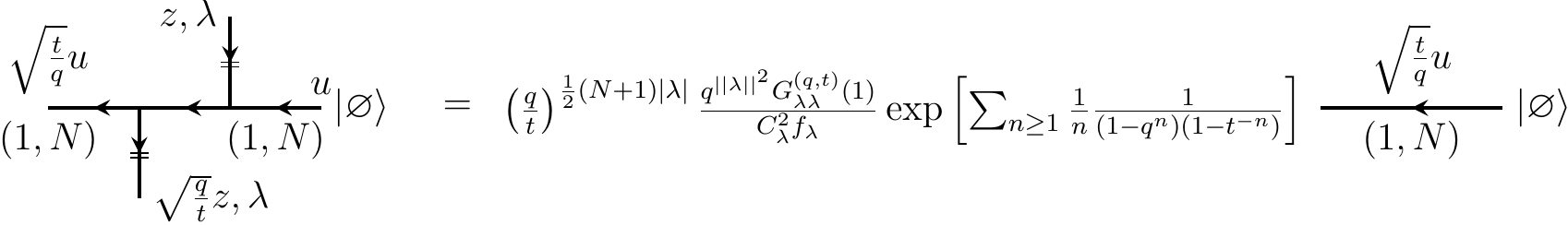}}\\
    \parbox{14cm}{\includegraphics[width=14cm]{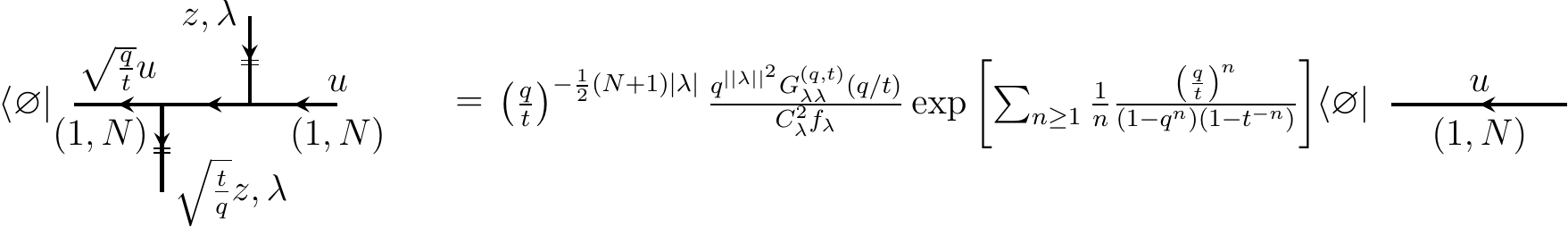}}\label{eq:75}
\end{gather}

Combining Eqs.~\eqref{eq:2} and~\eqref{eq:3} with Eq.~\eqref{eq:1}, one
finds that\footnote{One may find the identity:
$$  G_{\lambda \lambda}^{(q,t)}\left( \frac{q}{t} \right) = \left( \frac{q}{t} \right) ^{\vert \lambda \vert}
G_{\lambda \lambda}^{(q,t)} (1)
$$
useful in the derivation of Eq.~\eqref{eq:4}.}
\begin{framed}
\begin{multline}
  \label{eq:4}
 \Psi^{\lambda}\left(N,\sqrt{\frac{t}{q}} u\Big|\frac{q}{t} z\right)  =
  \exp \left[ \sum_{n
      \geq 1} \frac{1}{n}   \frac{1 - \left( \frac{q}{t}
      \right)^n}{(1-q^n)(1-t^{-n})} \right] \times\\
  \times
  \mathcal{T}^{\lambda}_{\lambda} \left(N+1, -u z \Big| \frac{q}{t} z ,
    \sqrt{\frac{q}{t}} z \right) \Psi^{\lambda} (N, u|z) \left( \mathcal{T}^{\lambda}_{\lambda} \left(N, u \Big| z ,
      \sqrt{\frac{q}{t}} z \right) \right)^{-1}
\end{multline}
\end{framed}
or, pictorially,
\begin{equation}
  \label{eq:58}
  \parbox{15cm}{\includegraphics[width=15cm]{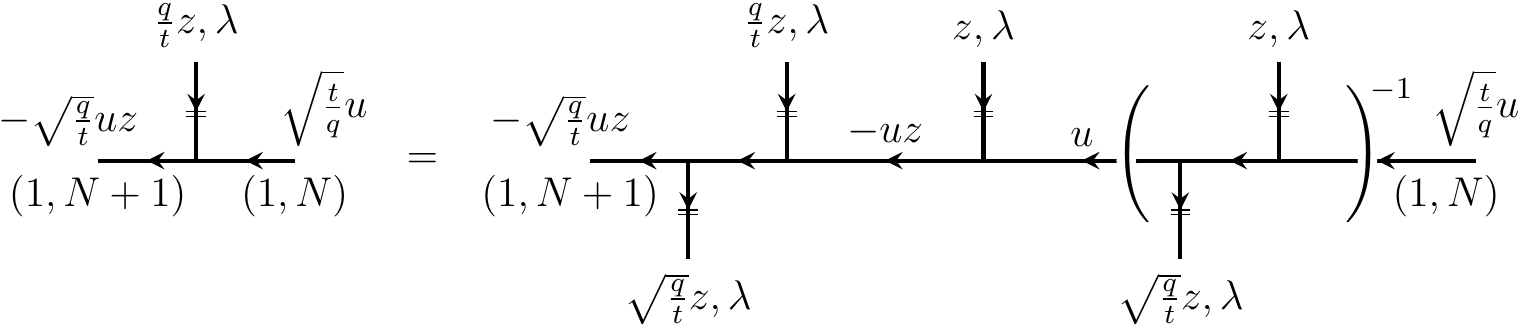}}
\end{equation}

We have thus expressed the shift operator in terms of the
$\mathcal{T}$-operators, algebraic objects with known
properties~\cite{NagoyaITEP2,NagoyaITEP3}. Notice that the r.h.s.\ of
Eq.~\eqref{eq:4} is automatically normal ordered since the left
(resp.\ right) $\mathcal{T}$-operator contains only the creation (resp.\
annihilation) operators. The horizontal spectral parameters of the
intertwiners are arranged so that they act consistently on the Fock
spaces. However, this is inessential, since the
$\mathcal{T}$-operators in the r.h.s.\ are actually independent of the
horizontal spectral parameter, as can be seen from
Eqs.~\eqref{eq:2},~\eqref{eq:3}.

The dual intertwiner $\Psi^{*}$ satisfies similar difference
equations:
\begin{equation}
  \label{eq:12}
\boxed{  \Psi^{*}_{\mu} \left(N, p u \Big|  w
  \right) =  p^{-|\mu|} \Psi^{*}_{\mu} \left(N, u \Big|  w
  \right)}
\end{equation}
\begin{framed}
  \begin{multline}
  \label{eq:5}
  \Psi^{*}_{\mu} \left(N, \sqrt{\frac{q}{t}} u \Big| \frac{q}{t} w \right) = \exp \left[ - \sum_{n
      \geq 1} \frac{1}{n}   \frac{1 - \left( \frac{q}{t}
      \right)^n}{(1-q^n)(1-t^{-n})} \right]\times\\
  \times
  \left( \mathcal{T}^{\mu}_{\mu} \left( N-1, - \sqrt{\frac{t}{q}} \frac{u}{w} \Big| \sqrt{\frac{q}{t}} w ,
      w \right) \right)^{-1} \Psi^{*}_{\mu} (N,u|w)
  \mathcal{T}^{\mu}_{\mu} \left( N,\sqrt{\frac{q}{t}} u \Big|
    \sqrt{\frac{q}{t}} w ,
    \frac{q}{t} w \right)
\end{multline}
\end{framed}

If one takes a product of \emph{several} intertwiners $\Psi$ and
$\Psi^{*}$ ending with the vacuum vectors on the both ends, after
shifting one of the vertical spectral parameters, one is able to
move the two $\mathcal{T}$-operators in Eq.~\eqref{eq:4} to the left
and to the right correspondingly, where they eventually annihilate the vacua. In the course of
doing this, one has to use the commutation relations between the
$\mathcal{T}$-operators and the intertwiners, which we describe in the
next section.

\subsection{Commutation of vertex and $\mathcal{T}$-operators}
\label{sec:commutation-vertex-t}
Let us prove the commutation relations which we use in the derivation
of the $(q,t)$-KZ equation. By direct computation, one gets
\begin{gather}
  \label{eq:16}
  \Psi^{*}_{\mu}(N+1,-uz|w) \Psi^{\lambda}(N,u|z) = \Psi^{\lambda}
  \left(N-1, - \frac{u}{w} \Big| z \right) \Psi^{*}_{\mu} (N,u|w)
  \Upsilon_{q,t} \left( \sqrt{\frac{q}{t}} \Big| \frac{z}{w}
  \right)\\
  \Psi^{\lambda_1}(N+1,-uz_2 | z_1) \Psi^{\lambda_2} (N,u|z_2) =
  \Psi^{\lambda_2}(N+1,-uz_1 | z_2) \Psi^{\lambda_1} (N,u|z_1)
  \mathcal{R}_{\lambda_2 \lambda_1} \left( \frac{z_2}{z_1} \right)
  \Upsilon_{q,t} \left( \frac{q}{t} \Big| \frac{z_1}{z_2}
  \right) \label{eq:17}\\
  \Psi^{*}_{\mu_1}\left(N-1,-\frac{u}{w_2}\Big| w_1 \right) \Psi^{*}_{\mu_2}(N,u|w_2) =
  \Psi^{*}_{\mu_2}\left(N-1, - \frac{u}{w_1} \Big| w_2 \right) \Psi^{*}_{\mu_1}(N,u|w_1)
  \frac{\Upsilon_{q,t}\left( 1 \Big| \frac{w_1}{w_2} \right)
  }{\mathcal{R}_{\mu_2 \mu_1}\left( \frac{w_2}{w_1}
    \right)} \label{eq:18}
\end{gather}
where
\begin{gather}
  \label{eq:9}
  \mathcal{R}_{\alpha \beta}(x) = \left( \frac{q}{t} \right)^{\frac{1}{2}(|\alpha|+|\beta|)}
  \frac{G_{\alpha \beta}^{(q,t)}(x)}{G_{\alpha \beta}^{(q,t)}\left(
      \frac{q}{t} x \right)}, \qquad \Upsilon_{q,t}( \alpha | x ) =
  \exp \left[ \sum_{n\geq 1} \frac{\alpha^n}{n} \frac{\left( x^n -
        x^{-n} \right)}{(1-q^n)(1-t^{-n})} \right]
\end{gather}
We see that
the commutation relations are in fact very simple. Apart from the
``anomalous factors'' $\Upsilon_{q,t}$, they are equivalent to two
copies of the Zamolodchikov algebra \cite{Zam} satisfied separately by $\Psi$ and
$\Psi^{*}$. The pictorial description of the commutation
relations~\eqref{eq:16}--\eqref{eq:18} is drawn as follows:
\begin{gather}
  \label{eq:61}
  \parbox{11cm}{\includegraphics[width=11cm]{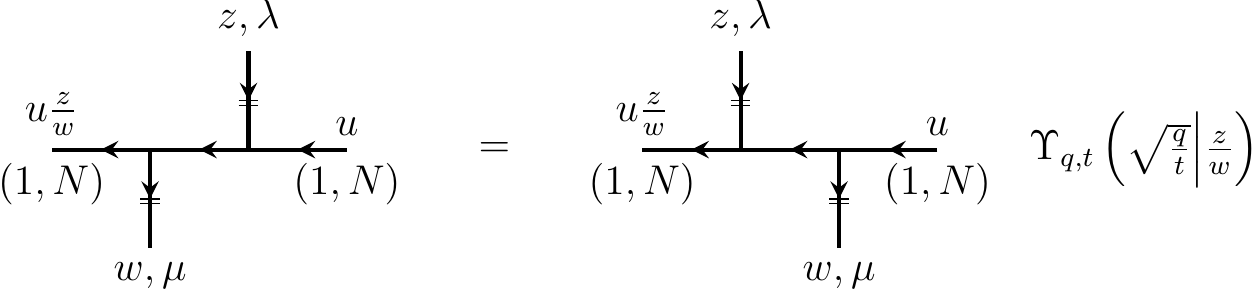}}\\
  \parbox{13cm}{\includegraphics[width=13cm]{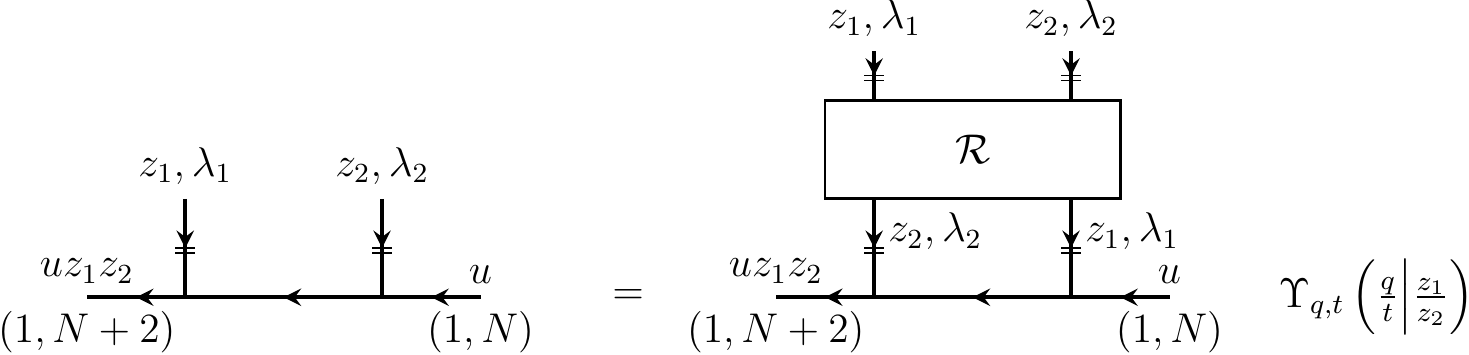}}\label{eq:62}\\
  \notag\\
      \parbox{13cm}{\includegraphics[width=13cm]{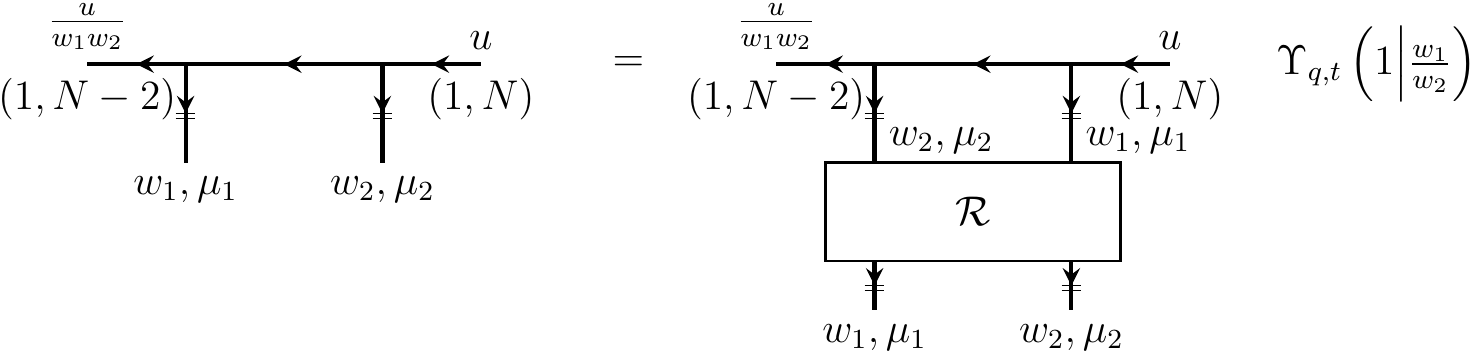}}\label{eq:63}
\end{gather}
Here the rectangular block denotes the $\mathcal{R}$-matrix in the
following convention:
\begin{equation}
  \label{eq:65}
  \parbox{7cm}{\includegraphics[width=7cm]{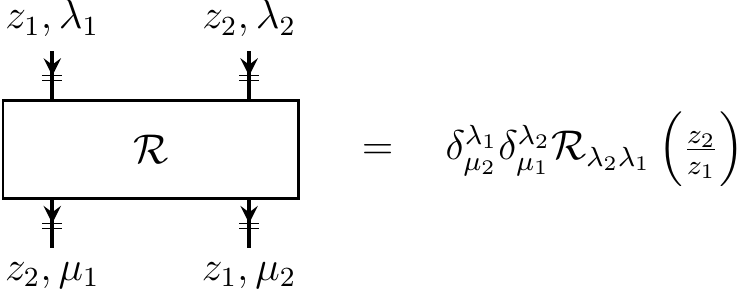}}
\end{equation}

Using Eqs.~\eqref{eq:16}--\eqref{eq:18}, one immediately gets the
commutation relations for the intertwiners $\Psi$, $\Psi^{*}$ and the
$\mathcal{T}$-operator:
\begin{framed}
\begin{gather}
  \label{eq:19}
  \mathcal{T}^{\mu}_{\nu}(N+1,-uy| z,w) \Psi^{\lambda}(N,u|y) =
  \mathcal{R}_{\lambda \mu} \left( \frac{y}{z} \right)
  \Psi^{\lambda}\left(N,u \frac{z}{w} \Big| y \right)
  \mathcal{T}^{\mu}_{\nu}(N,u|z,w) \Upsilon_{q,t}\left( \frac{q}{t}
    \Big| \frac{z}{y} \right)\Upsilon_{q,t}\left(
    \sqrt{\frac{q}{t}} \Big| \frac{y}{w} \right)\\
  \mathcal{T}^{\mu}_{\nu}\left(N-1, -\frac{u}{y}\Big|z, w \right) \Psi^{*}_{\lambda}(N,u|y) =
  \frac{1}{\mathcal{R}_{\lambda \nu} \left( \frac{y}{w} \right)}
  \Psi^{*}_{\lambda}\left(N, u \frac{z}{w}\Big| y\right)
  \mathcal{T}^{\mu}_{\nu}(N,u|z,w) \Upsilon_{q,t}\left( 1 \Big| \frac{w}{y}
  \right)\Upsilon_{q,t}\left( \sqrt{\frac{q}{t}} \Big| \frac{y}{z}
  \right)
  \label{eq:74}
\end{gather}
\end{framed}

\subsection{$(q,t)$-KZ equation for DIM}
\label{sec:q-kz-equation}
Let us introduce a compact notation for the vacuum correlation matrix
element of the product of $\Psi$ and $\Psi^{*}$ intertwiners:
\begin{equation}
  \label{eq:15}
    \mathcal{G}_{\mu_1 \cdots \mu_m }^{\lambda_1 \cdots \lambda_n} (u_n|
  \begin{smallmatrix}
    z_1, \ldots , z_n\\
    w_1, \ldots , w_m
  \end{smallmatrix}
)
\stackrel{\mathrm{def}}{=} \langle \varnothing | \Psi^{*}_{\mu_1}(n-m+1, v_1|w_1) \cdots
\Psi^{*}_{\mu_m} (n, v_m | w_m) \Psi^{\lambda_1}\left(n-1,  u_1
    | z_1 \right) \cdots \Psi^{\lambda_n}\left( 0, u_n |
    z_n \right) |
  \varnothing\rangle
\end{equation}
One can draw $\mathcal{G}$ as follows:
\begin{equation}
  \label{eq:59}
    \parbox{10cm}{\includegraphics[width=10cm]{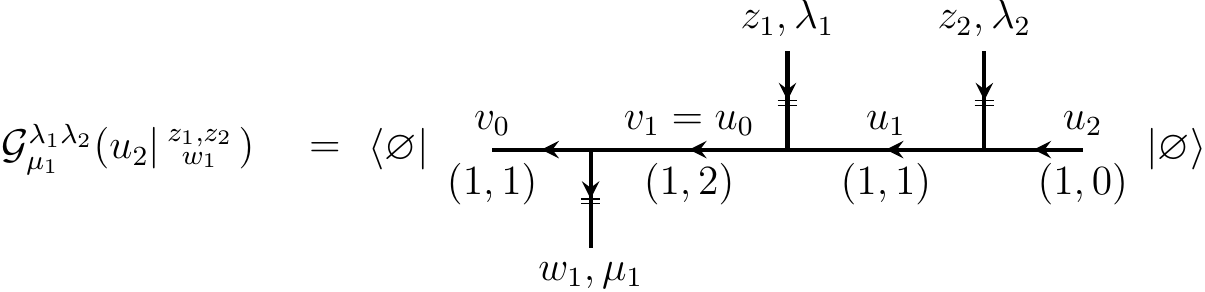}}
\end{equation}

Here the only independent horizontal spectral parameter is $u_n$ and
all other horizontal spectral parameters are determined by the
relations
\begin{gather}
  u_i = -z_{i+1} u_{i+1}, \qquad i=1\ldots (n-1),\notag\\
  v_m = - z_1 u_1, \label{eq:20}
  \\
  v_j = - \frac{v_{j+1}}{w_{j+1}} \qquad j=1\ldots (m-1).\notag
\end{gather}
Without loss of generality, we chose the slope of the rightmost
horizontal leg to be $(1,0)$. General slopes of the form $(1,N)$ can
be obtained by the application of the automorphism $\mathfrak{T} = \left(
  \begin{smallmatrix}
    1 & 1\\
    0 & 1
  \end{smallmatrix}
\right) \in SL(2, \mathbb{Z})$ of DIM. This automorphism transforms
the horizontal representations changing their slopes in the natural
way, $(n_1,n_2) \stackrel{\mathfrak{T}}{\mapsto} (n_1,n_1 + n_2) $. It acts
diagonally on the vertical representation with matrix elements in the
Macdonald basis given by:
\begin{equation}
  \label{eq:22}
  \mathfrak{T}^{\lambda}_{\mu} = \left\langle M_{\lambda},z|\mathfrak{T}|M_{\mu},z \right\rangle
  = \delta_{\lambda\mu} (-z)^{|\lambda|} f_{\lambda}
\end{equation}
Due to the intertwining property of $\Psi$ and $\Psi^{*}$, one can
rewrite the change of slope of the horizontal legs in terms of the
action of $T$ on the vertical leg:
\begin{gather}
  \label{eq:21}
  \Psi^{\lambda} (N+1, u|z) = \mathfrak{T}^{-1} \Psi^{\lambda} (N, u|z) \mathfrak{T} =
  \sum_{\mu} (\mathfrak{T}^{-1})^{\lambda}_{\mu}\Psi^{\mu} (N, u|z)=
  (-z)^{-|\lambda|} f_{\lambda}^{-1} \Psi^{\lambda} (N, u|z)\\
  \Psi^{*}_{\mu} (N+1, u|w) = \mathfrak{T}^{-1} \Psi^{*}_{\mu} (N, u|w) \mathfrak{T} =
  \sum_{\nu} \mathfrak{T}^{\nu}_{\mu} \Psi^{*}_{\nu} (N, u|w) = (-w)^{|\mu|}
  f_{\mu} \Psi^{*}_{\mu} (N, u|w).
\end{gather}
Therefore, changing the overall slope in $\mathcal{G}$ amounts to the
multiplication by certain simple factors.

Let us act on the correlator $\mathcal{G}$ with the difference
operator $\left( \frac{q}{t} \right)^{z_k \partial_{z_k}}$. According
to Eq.~\eqref{eq:11}, this creates a pair of $\mathcal{T}$-operators
surrounding the intertwiner $\Psi$. We then push the
$\mathcal{T}$-operators to the ends of the correlator using the
commutation relations,~\eqref{eq:19},~(\ref{eq:74}). Eventually, the
$\mathcal{T}$-operators act on the vacuum giving the prefactor, which
cancels the prefactor in~(\ref{eq:4}) and leaving behind
the original correlator $\mathcal{G}$.

To write down the resulting equation, it is convenient to introduce the
modified $\mathcal{R}$-matrix $\widetilde{\mathcal{R}}_{\lambda
  \mu}(x)$ differing from $\mathcal{R}_{\lambda \mu}(x)$ by a scalar
factor\footnote{In the notation of~\cite{FJMM}, our
  $\mathcal{R}$-matrix $\mathcal{R}_{\lambda\mu}(x)$ is denoted by
  $\bar{R}_{\mathcal{F}, \mathcal{F}}(x)$ and the $\mathcal{R}$-matrix
  $\widetilde{\mathcal{R}}_{\lambda \mu}(x)$ with the additional scalar
  factor introduced in Eq.~\eqref{eq:8} is denoted simply by
  $\mathcal{R}(x)$.}:
\begin{equation}
  \label{eq:8}
  \widetilde{\mathcal{R}}_{\lambda \mu}(x) \stackrel{\mathrm{def}}{=} \mathcal{R}_{\lambda
    \mu}(x) \frac{\Upsilon_{q,t}(1\vert x)}
{\Upsilon_{q,t}(\sqrt{\frac{q}{t}} \vert\sqrt{\frac{t}{q}}  x)} = \mathcal{R}_{\lambda \mu}(x)
  \exp \left[- \sum_{n \geq 1} \frac{x^{-n}}{n} \frac{1-
      \left( \frac{q}{t} \right)^n}{(1-q^n)(1-t^{-n})} \right]\end{equation}
The $(q,t)$-KZ equation for DIM intertwiners is then written as follows:
\begin{framed}
  \begin{multline}
  \label{eq:30}
  \left( \frac{q}{t} \right)^{z_k \partial_{z_k} - \frac{1}{2} u_n \partial_{u_n}} \mathcal{G}_{\mu_1 \cdots \mu_m }^{\lambda_1 \cdots \lambda_n} (u_n|
  \begin{smallmatrix}
    z_1, \ldots , z_n\\
    w_1, \ldots , w_m
  \end{smallmatrix}
  ) =\\
  =\left( \frac{q}{t} \right)^{\frac{m-n}{2}|\lambda_k|} \prod_{l=1}^m
  \widetilde{\mathcal{R}}_{\mu_l \lambda_k} \left( \sqrt{\frac{t}{q}}
    \frac{w_l}{z_k} \right) \prod_{i < k}
  \frac{1}{\widetilde{\mathcal{R}}_{\lambda_i \lambda_k} \left(
      \frac{t z_i}{q z_k} \right)} \prod_{j > k}
  \widetilde{\mathcal{R}}_{\lambda_k \lambda_j} \left( \frac{z_k}{z_j}
  \right) \mathcal{G}_{\mu_1 \cdots \mu_m }^{\lambda_1 \cdots
    \lambda_n} (u_n|
  \begin{smallmatrix}
    z_1, \ldots , z_n\\
    w_1, \ldots , w_m
  \end{smallmatrix}
  )
\end{multline}
\end{framed}
\noindent
Here the extra shift of $u_n$ by $\sqrt{\frac{t}{q}}$ was introduced
to conform with Eq.~\eqref{eq:11}. A similar equation holds for the
shift in the vertical spectral parameters of the dual intertwiners
$\Psi^{*}$:
\begin{framed}
  \begin{multline}
  \label{eq:31}
  \left( \frac{q}{t} \right)^{w_k \partial_{w_k} + \frac{1}{2}
    u_n \partial_{u_n}} \mathcal{G}_{\mu_1 \cdots \mu_m }^{\lambda_1
    \cdots \lambda_n} (u_n|
  \begin{smallmatrix}
    z_1, \ldots , z_n\\
    w_1, \ldots , w_m
  \end{smallmatrix}
  ) =\\
  =\left( \frac{q}{t} \right)^{\frac{n-m}{2}|\mu_k|} \prod_{i < k}
  \frac{1}{\widetilde{\mathcal{R}}_{\mu_i \mu_k} \left(
      \frac{w_i}{w_k} \right)} \prod_{j > k}
  \widetilde{\mathcal{R}}_{\mu_k \mu_j} \left( \frac{qw_k}{tw_j}
  \right) \prod_{l=1}^n
  \frac{1}{\widetilde{\mathcal{R}}_{\mu_k \lambda_l} \left( \sqrt{\frac{q}{t}}
    \frac{w_k}{z_l} \right)}  \mathcal{G}_{\mu_1 \cdots \mu_m }^{\lambda_1 \cdots
    \lambda_n} (u_n|
  \begin{smallmatrix}
    z_1, \ldots , z_n\\
    w_1, \ldots , w_m
  \end{smallmatrix}
  )
\end{multline}
\end{framed}
\noindent
These are the main results of this section. We will now give explicit
solutions to these equations, which are equal to refined topological
amplitudes on the strip geometry.

\subsection{Explicit solution to DIM $(q,t)$-KZ}
\label{sec:solution-q-kz}

As we stressed earlier, the DIM algebra $U_{q,t}(\widehat{\widehat{\mathfrak{gl}}}_1)$ is quasi-Abelian, which means that there are no screening charges in this case. Indeed, as we explained in \cite{NagoyaITEP}, the screening charges are associated with the network models containing two ($q$-Virasoro) or more ($q$-$W$-algebra) horizontal lines. In fact, the number of horizontal lines corresponds to the first central charge: if it is equal to an integer $k$, we have $q$-$W^{(k)}$-algebra. In particular, the deformation of WZW model is described by one horizontal line (see, e.g., the figure in (\ref{eq:59}) or Fig.1). Thus, in this case, there are no screening charges, and the ``Abelian'' $(q,t)$-KZ equation
has a solution in terms of products of propagators. In our case, the
role of the propagators is played by the functions
\begin{equation}
  \label{eq:32}
  F_{\lambda \mu}(x) \stackrel{\mathrm{def}}{=} x^{-|\mu|} G_{\lambda \mu}(x) \exp \left[
    \sum_{n \geq 1} \frac{1}{n} \left( \frac{t}{q} x \right)^{-n} \frac{1}{(1-q^n)(1-t^{-n})} \right].
\end{equation}
Notice that the exponential factor is nothing but the generating
function of equivariant Euler characters of the Hilbert scheme
$\mathrm{Hilb}(\mathbb{C}^2)$,~\cite{Nakajima:2003pg} or the refined
topological string amplitude of the resolved conifold.

The function $F_{\lambda \mu}(x)$ satisfies the following fundamental
difference equation:
\begin{equation}
  \label{eq:57}
  \frac{F_{\lambda \mu}\left( \frac{q}{t} x \right)}{F_{\lambda
      \mu}(x)} = \left( \frac{q}{t} \right)^{\frac{|\lambda|-|\mu|}{2}} \frac{1}{\widetilde{\mathcal{R}}_{\lambda \mu}(x)}.
\end{equation}
In terms of the function $F_{\lambda \mu}(x)$, the solution reads as follows:
\begin{equation}
  \label{eq:14}
  \mathcal{G}_{\mu_1 \cdots \mu_m }^{\lambda_1 \cdots \lambda_n} (u_n|
  \begin{smallmatrix}
    z_1, \ldots , z_n\\
    w_1, \ldots , w_m
  \end{smallmatrix}
) =
  u_n^{\sum_{i=1}^n |\lambda_i| - \sum_{j=1}^m
    |\mu_j|} \frac{\prod_{k=1}^m \prod_{i=1}^n F_{\mu_k
      \lambda_i} \left( \sqrt{\frac{q}{t}} \frac{w_k}{z_i}
      \right)}{\prod_{k<l}^m F_{\mu_k \mu_l} \left( \frac{q w_k}{t
        w_l} \right) \prod_{i<j}^n F_{\lambda_i \lambda_j} \left( \frac{z_i}{z_j} \right)}
\end{equation}
The solution is of course defined up to multiplication by a
$\left(\frac{q}{t}\right)$-periodic function of the spectral
parameters $z_i$ and $w_j$. It is curious to notice that the form of
the solution resembles the free fermion correlator.

The structure on the r.h.s.\ of Eq.~\eqref{eq:14} is consistent with
the fact that by the spectral duality \cite{spec-dual} it gives a building block of the
Nekrasov partition function for $\mathcal{N} =2$ linear quiver gauge
theory. That is, a pair of edges on the opposite sides of the
(horizontal) line gives the bifundamental matter contribution, while a
pair on the same side gives the gauge (root) boson contribution.

The braiding transformations $\hat{\mathcal{B}}^{i,i+1}$ and
$\hat{\mathcal{B}}_{j,j+1}$ (permuting the positions of $\Psi$ and
$\Psi^{*}$ intertwiners respectively) of the solution are defined as
follows:
\begin{gather}
  \label{eq:33}
  \hat{\mathcal{B}}^{i,i+1} \mathcal{G}_{\mu_1 \cdots \mu_m }^{\lambda_1 \cdots \lambda_n} (u_n|
  \begin{smallmatrix}
    z_1, \ldots , z_n\\
    w_1, \ldots , w_m
  \end{smallmatrix}
) \stackrel{\mathrm{def}}{=}  \mathcal{G}_{\mu_1 \cdots \mu_m
}^{\lambda_1 \cdots \lambda_{i+1} \lambda_i \cdots \lambda_n} (u_n|
  \begin{smallmatrix}
    z_1, \ldots z_{i+1}, z_i,\ldots , z_n\\
    w_1, \ldots , w_m
  \end{smallmatrix}
  ),\\
  \hat{\mathcal{B}}_{j,j+1} \mathcal{G}_{\mu_1 \cdots \mu_m
  }^{\lambda_1 \cdots \lambda_n} (u_n|
  \begin{smallmatrix}
    z_1, \ldots , z_n\\
    w_1, \ldots , w_m
  \end{smallmatrix}
) \stackrel{\mathrm{def}}{=}  \mathcal{G}_{\mu_1 \cdots, \mu_{j+1}, \mu_j ,\cdots \mu_m
}^{\lambda_1 \cdots  \lambda_n} (u_n|
  \begin{smallmatrix}
    z_1, \ldots , z_n\\
    w_1, \ldots, w_{j+1}, w_j, \ldots , w_m
  \end{smallmatrix}
  )
\end{gather}
Evidently, the two sets of operators commute,
$[\hat{\mathcal{B}}^{i,i+1}, \hat{\mathcal{B}}_{i,i+1}] = 0$ and
separately satisfy the braid group relations. The explicit computation
shows that the braiding is performed by the DIM $\mathcal{R}$-matrix
with the anomalous factor as in Eqs.~\eqref{eq:17},~\eqref{eq:18}:
\begin{gather}
  \label{eq:34}
  \hat{\mathcal{B}}^{i,i+1} \mathcal{G}_{\mu_1 \cdots \mu_m }^{\lambda_1 \cdots \lambda_n} (u_n|
  \begin{smallmatrix}
    z_1, \ldots , z_n\\
    w_1, \ldots , w_m
  \end{smallmatrix}
) =  \frac{f_{\lambda_i}}{f_{\lambda_{i+1}}} \mathcal{R}_{\lambda_i \lambda_{i+1}}
\left( \frac{z_i}{z_{i+1}} \right) \Upsilon_{q,t} \left(
  \frac{q}{t} \Big| \frac{z_{i+1}}{z_i} \right) \mathcal{G}_{\mu_1 \cdots \mu_m
}^{\lambda_1 \cdots  \lambda_n} (u_n|
  \begin{smallmatrix}
    z_1, \ldots , z_n\\
    w_1, \ldots , w_m
  \end{smallmatrix}
  ),\\
\label{eq:6}
\hat{\mathcal{B}}_{j,j+1} \mathcal{G}_{\mu_1 \cdots \mu_m }^{\lambda_1
  \cdots \lambda_n} (u_n|
  \begin{smallmatrix}
    z_1, \ldots , z_n\\
    w_1, \ldots , w_m
  \end{smallmatrix}
  ) = \left( \frac{q}{t} \right)^{|\mu_j|-|\mu_{j+1}|}
  \frac{f_{\mu_j}}{f_{\mu_{j+1}}} \frac{\Upsilon_{q,t} \left( 1 \Big| \frac{w_{j+1}}{w_j} \right)}{\mathcal{R}_{\mu_j
      \mu_{j+1}} \left( \frac{w_j}{w_{j+1}} \right)}
  \mathcal{G}_{\mu_1 \cdots \mu_m }^{\lambda_1 \cdots  \lambda_n} (u_n|
  \begin{smallmatrix}
    z_1, \ldots , z_n\\
    w_1, \ldots , w_m
  \end{smallmatrix}
  ).
\end{gather}
Extra constant factors are not captured by our solution (which is
defined up to a $\left( \frac{q}{t} \right)$-periodic function) and
appear in the r.h.s.\ of Eqs.~\eqref{eq:34},~\eqref{eq:6}. They can be
eliminated by a simple change of the normalization of solutions.

\section{Elliptic $(q,t)$-KZ equation for DIM}
\label{sec:elliptic-q-kz}

Using the difference operator from sec.~\ref{sec:difference-operator}
and commutation relations from sec.~\ref{sec:commutation-vertex-t},
one can also derive \emph{elliptic} $(q,t)$-KZ equation for the
\emph{trace} of a product of intertwiners, which is a DIM counterpart
of the KZBE on torus (see a similar equation for the quantum affine
algebra in \cite{Etin}). As in the previous section, we first sketch
the main idea of the derivation and then give the precise expressions
with correct arguments and prefactors.

We start with the trace of the following form
\begin{equation}
  \label{eq:10}
  ``\Tr \left( Q^d Q_{\perp}^{d_{\perp}} \Psi \cdots  \Psi \right)"
\end{equation}
Here we added the grading operators $d$ and $d_{\perp}$ for the sake
of generality. In a moment, we will see that they are in fact customary
to get a meaningful equation for the trace. Proceeding as in the
previous section, we rewrite the shift operator in terms of the
$\mathcal{T}$-operators as sketched in Eq.~\eqref{eq:25} and move the
$\mathcal{T}$-operators to the left and right of $\Psi$. This gives a
product of $\mathcal{R}$-matrices as in Eq.~\eqref{eq:27} of the
previous section:
\begin{multline}
  \label{eq:7}
  ``\Tr \left( Q^d Q_{\perp}^{d_{\perp}} \Psi \cdots \Psi \left( \frac{q}{t} z \right)
  \cdots \Psi \right) = \Tr \left( Q^d Q_{\perp}^{d_{\perp}} \Psi
  \cdots \mathcal{T}(z) \Psi (z) \mathcal{T}^{-1}(z)
  \cdots \Psi \right)
   =\\
=\left(\prod \mathcal{R} \right) \Tr \left( Q^d Q_{\perp}^{d_{\perp}} \mathcal{T}(z) \Psi
  \cdots  \Psi (z)
  \cdots \Psi \mathcal{T}^{-1}(z) \right)"
\end{multline}
However, now the $\mathcal{T}$-operators act not on the vacuum vectors but
under the trace. Therefore, using the cyclic property of the trace,
we move them further \emph{through each other} so that they reappear
on the opposite side of the expression. Taking a full circle, we obtain
the trace which looks similar to the initial one with the arguments of
the $\mathcal{T}$-operators shifted by $Q$ (note that the products of $\mathcal{R}$-matrices in (\ref{eq:7}) and (\ref{eq:13}) are different):
\begin{equation}
  \label{eq:13}
    ``\Tr \left( Q^d Q_{\perp}^{d_{\perp}} \Psi \cdots \Psi \left( \frac{q}{t} z \right)
  \cdots \Psi \right) =  \left(\prod \mathcal{R} \right) \Tr \left( Q^d Q_{\perp}^{d_{\perp}} \Psi
  \cdots \mathcal{T}(Qz) \Psi (z) \mathcal{T}^{-1}(Q^{-1}z)
  \cdots \Psi \right)"
\end{equation}
We repeat the cyclic transfer of the $\mathcal{T}$-operators so that
their arguments are shifted by higher and higher powers of $Q$. At the
same time, one gets more and more $\mathcal{R}$-matrices at the r.h.s. of
Eq.~\eqref{eq:13}. The crucial point is that the two
$\mathcal{T}$-operators featuring in Eq.~\eqref{eq:13} are
\emph{Taylor} (not \emph{Laurent}) series in $Q$ and $Q^{-1}$
respectively. If we assume that $|Q|<1$, then for high enough positive
powers of $Q$ the $\mathcal{T}$-operators converge to a \emph{scalar
  factor.} After that we are left with an \emph{infinite} product of
$\mathcal{R}$-matrices, which can be packed into the \emph{finite}
product of the new \emph{elliptic} $\mathcal{R}$-matrices which we
denote by the gothic letter $\mathfrak{R}$:
\begin{equation}
  \label{eq:35}
  ``\Tr \left( Q^d Q_{\perp}^{d_{\perp}} \Psi \cdots \Psi \left( \frac{q}{t} z \right)
    \cdots \Psi \right) =  \left(\prod \mathfrak{R} \right) \Tr \left( Q^d Q_{\perp}^{d_{\perp}} \Psi
    \cdots  \Psi (z)
    \cdots \Psi \right)"
\end{equation}
This equation is the DIM version of the elliptic $(q,t)$-KZ equation
usually satisfied by torus blocks of the affine algebra. The procedure itself is by essence the "averaging" procedure of constructing elliptic ${\cal R}$-matrices \cite{RF}. Let
us now proceed to the detailed derivation.

\subsection{Matching the Fock spaces}
\label{sec:matching-fock-spaces}
To take a trace of a product of intertwiners, one has to ensure that
this product is an endomorphism of some Fock space, i.e.\ that the
Fock spaces on the incoming and outgoing horizontal legs coincide.

First of all, the slopes should match. In the previous section, we
adopted the convention that the rightmost intertwiner acts on the space
with slope $(1,0)$. After $n$ $\Psi$ intertwiners and $m$ $\Psi^{*}$
acted on the space, the slope changes to $(1,n-m)$. We therefore
conclude that under the trace the number of $\Psi$ should be equal to
the number of $\Psi^{*}$, i.e.\ exactly the case of balanced network,
$n=m$.

Secondly, the spectral parameters on the left and right should
coincide. Let the rightmost intertwiner act on the Fock space with the
horizontal spectral parameter $u_n$. Then, using relations~\eqref{eq:20}, we obtain the horizontal spectral parameter
$v_0$ to the left of all the intertwiners:
\begin{equation}
  \label{eq:37}
  v_0 = u_n \prod_{i=1}^n\frac{z_i}{w_i}.
\end{equation}
Notice, however, that the operator $Q_{\perp}^{d_{\perp}}$
\emph{shifts} the horizontal spectral parameter from $v_0$ to
$Q_{\perp} v_0$. To be able to take the trace we, therefore, will
henceforth set
\begin{equation}
  \label{eq:38}
\boxed{  Q_{\perp} = \prod_{i=1}^n\frac{w_i}{z_i}}
\end{equation}
in all expressions. As in the previous section, we introduce the
notation for the trace of intertwiners:
\begin{equation}
  \label{eq:39}
  \mathfrak{G}_{\mu_1 \cdots \mu_n}^{\lambda_1 \cdots \lambda_n} (u_n|Q|
  \begin{smallmatrix}
    z_1, \ldots , z_n\\
    w_1, \ldots , w_n
  \end{smallmatrix}
  )
  \stackrel{\mathrm{def}}{=} \Tr\left\{ Q^d Q_{\perp}^{d_{\perp}} \Psi^{*}_{\mu_1}(1, v_1|w_1) \cdots
  \Psi^{*}_{\mu_n} (n, v_n | w_n) \Psi^{\lambda_1}\left(n-1,  u_1
    | z_1 \right) \cdots \Psi^{\lambda_n}\left( 0, u_n |
    z_n \right) \right\}
\end{equation}
We can draw the corresponding picture denoting the trace pairing as
wavy lines and the grading operator insertions as boxes:
\begin{equation}
  \label{eq:60}
      \parbox{13cm}{\includegraphics[width=13cm]{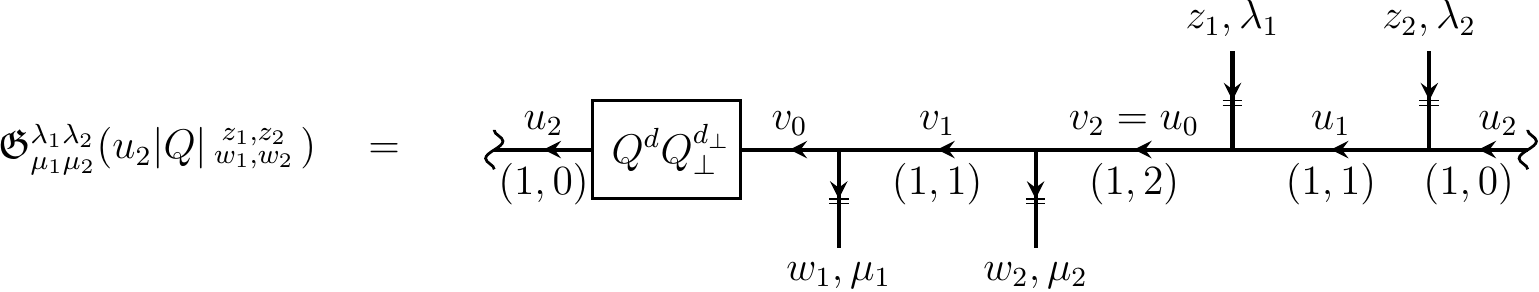}}
\end{equation}
The parameter $Q$ remains free and can be interpreted as the
exponentiated modular parameter of the torus on which the conformal
block of DIM lives. The results of the previous section are recovered
in the limit $Q \to 0$, when the torus degenerates into a cylinder.

\subsection{$q$-KZ equation for the trace of intertwiners}
\label{sec:elliptic-q-kz-1}
The derivation of the $q$-KZ equation for the trace of intertwiners
goes along the same lines as that of the equation for the vacuum
matrix element. We express the shift operator through the
$\mathcal{T}$-operators using Eq.~\eqref{eq:4}. We then use the
commutation relations for the intertwiners and
$\mathcal{T}$-operators~\eqref{eq:19} to move the
$\mathcal{T}$-operators cyclically under the trace. After each cycle
the vertical spectral parameters of the $\mathcal{T}$-operators are
multiplied by $Q$:
\begin{equation}
  \label{eq:40}
  Q^d\mathcal{T}^{\lambda}_{\lambda}(N,u|z,w) =
  \mathcal{T}^{\lambda}_{\lambda}(N,u|Qz,Qw) Q^d
\end{equation}
From Eqs.~\eqref{eq:2}--\eqref{eq:3} assuming $|Q|<1$ in the limit of
large number $k$ of cycles, one has
\begin{gather}
  \mathcal{T}^{\lambda}_{\lambda} \left(N, u \Big| Q^{-k} z ,
    \sqrt{\frac{q}{t}} Q^{-k} z \right) \xrightarrow[k \to \infty]{}
  \left( \frac{q}{t} \right)^{\frac{1}{2} (N+1) \vert\lambda\vert}
\frac{q^{|| \lambda ||^2} G_{\lambda\lambda}^{(q,t)}(1)}{C_\lambda^2
  f_\lambda} \exp \left[ \sum_{n \geq 1}
    \frac{1}{n} \frac{1}{(1-q^n)(1-t^{-n})} \right]\label{eq:41}
  \\
  \notag\\
  \mathcal{T}^{\lambda}_{\lambda} \left( N, u \Big| Q^k z ,
    \sqrt{\frac{t}{q}} Q^k z \right) \xrightarrow[k \to \infty]{}
  \left( \frac{q}{t} \right)^{- \frac{1}{2} (N+1) \vert\lambda\vert}
\frac{q^{|| \lambda ||^2} G_{\lambda\lambda}^{(q,t)}(q/t)}{C_\lambda^2 f_\lambda}
 \exp \left[ \sum_{n \geq 1}
    \frac{1}{n} \frac{\left( \frac{q}{t} \right)^n}{(1-q^n)(1-t^{-n})}
  \right]\label{eq:42}
\end{gather}
Thus, after making a sufficient number of cycles, the
$\mathcal{T}$-operators essentially disappear and we are left with the
infinite number of $\mathcal{R}$-matrices. We pack the infinite
product into the combination of \emph{three} different elliptic
$\mathcal{R}$-matrices differing by slight shifts of arguments:
\begin{gather}
  \widetilde{\mathfrak{R}}^{(1)}_{\lambda \mu} (x|Q)
  \stackrel{\mathrm{def}}{=} \prod_{k \geq 0}
  \frac{\mathcal{R}_{\lambda \mu}(x
    Q^{k+1})}{\mathcal{R}_{\mu \lambda}(x^{-1} Q^k)} \exp
  \left[ \sum_{n \geq 1} \frac{1-\left( \frac{q}{t} \right)^n}{n}
    \frac{(Q x)^n - x^{-n}}{(1-q^n)(1-t^{-n})(1-Q^n)}  \right],\notag\\
  \widetilde{\mathfrak{R}}^{(2)}_{\lambda \mu} (x|Q)
  \stackrel{\mathrm{def}}{=} \prod_{k \geq 0}
  \frac{\mathcal{R}_{\lambda \mu}\left( \frac{q}{t} x
      Q^{k+1}\right)}{\mathcal{R}_{\mu \lambda}(x^{-1}
    Q^k)} \exp \left[ \sum_{n \geq 1} \frac{1-\left( \frac{q}{t}
      \right)^n}{n} \frac{ \left( \frac{q}{t} Q x\right)^n -
      x^{-n}}{(1-q^n)(1-t^{-n})(1-Q^n)} \right], \label{eq:44}
  \\
  \widetilde{\mathfrak{R}}^{(3)}_{\lambda \mu} (x|Q)
  \stackrel{\mathrm{def}}{=} \prod_{k \geq 0}
  \frac{\mathcal{R}_{\lambda \mu}\left( \frac{t}{q} x
      Q^{k+1}\right)}{\mathcal{R}_{\mu \lambda}(x^{-1}
    Q^k)} \exp \left[ \sum_{n \geq 1} \frac{1-\left( \frac{q}{t}
      \right)^n}{n} \frac{ \left( \frac{t}{q} Q x\right)^n -
      x^{-n}}{(1-q^n)(1-t^{-n})(1-Q^n)} \right],\notag
\end{gather}
These $\mathcal{R}$-matrices should be associated with the elliptic
DIM algebra~\cite{ellDIM}. Eventually, we obtain elliptic KZ equations
for the shifts of $\Psi$ and $\Psi^{*}$ intertwiners:
\begin{framed}
  \begin{multline}
    \label{eq:46}
    \left( \frac{q}{t} \right)^{z_k \partial_{z_k} - \frac{1}{2}
      u_n \partial_{u_n}} \mathfrak{G}_{\mu_1 \cdots \mu_n
    }^{\lambda_1 \cdots \lambda_n} (u_n|Q|
  \begin{smallmatrix}
    z_1, \ldots , z_n\\
    w_1, \ldots , w_n
  \end{smallmatrix}
  ) =\\
  =\prod_{l=1}^n \widetilde{\mathfrak{R}}^{(1)}_{\mu_l \lambda_k} \left(
    \sqrt{\frac{t}{q}} \frac{w_l}{z_k} \Big| Q \right) \prod_{i < k}
  \frac{1}{\widetilde{\mathfrak{R}}^{(2)}_{\lambda_i \lambda_k} \left(
      \frac{t z_i}{q z_k} \Big| Q \right)} \prod_{j > k}
  \widetilde{\mathfrak{R}}^{(2)}_{\lambda_k \lambda_j} \left(
    \frac{z_k}{z_j} \Big| Q \right) \mathfrak{G}_{\mu_1 \cdots \mu_n
  }^{\lambda_1 \cdots \lambda_n} (u_n|Q|
  \begin{smallmatrix}
    z_1, \ldots , z_n\\
    w_1, \ldots , w_n
  \end{smallmatrix}
  )
\end{multline}
  \begin{multline}
\label{eq:45}  \left( \frac{q}{t} \right)^{w_k \partial_{w_k} + \frac{1}{2}
    u_n \partial_{u_n}} \mathfrak{G}_{\mu_1 \cdots \mu_n }^{\lambda_1
    \cdots \lambda_n} (u_n|Q|
  \begin{smallmatrix}
    z_1, \ldots , z_n\\
    w_1, \ldots , w_n
  \end{smallmatrix}
  ) =\\
  = \prod_{i < k} \frac{1}{\widetilde{\mathfrak{R}}^{(3)}_{\mu_i
      \mu_k} \left( \frac{w_i}{w_k} \Big| Q \right)} \prod_{j > k}
  \widetilde{\mathfrak{R}}^{(3)}_{\mu_k \mu_j} \left(
    \frac{qw_k}{tw_j} \Big| Q \right) \prod_{l=1}^n
  \frac{1}{\widetilde{\mathfrak{R}}^{(1)}_{\mu_k \lambda_l} \left(
      \sqrt{\frac{q}{t}} \frac{w_k}{z_l}\Big| Q \right)}
  \mathcal{G}_{\mu_1 \cdots \mu_n}^{\lambda_1 \cdots \lambda_n}
  (u_n|Q|
  \begin{smallmatrix}
    z_1, \ldots , z_n\\
    w_1, \ldots , w_n
  \end{smallmatrix}
  )
\end{multline}
\end{framed}
\noindent
The form of KZBE~\eqref{eq:46},~\eqref{eq:45} is very similar to that
of the KZE~\eqref{eq:30},~\eqref{eq:31}, the only important difference
being the elliptic $\mathcal{R}$-matrices. In the next subsection, we
find the explicit solution to the elliptic KZ equations.

\subsection{Explicit solution for elliptic $(q,t)$-KZ equation}
\label{sec:expl-solut-ellipt}
Proceeding as in the case of $(q,t)$-KZE, we introduce three propagators
corresponding to the three $\mathcal{R}$-matrices~\eqref{eq:44} (again the
difference is in slight shifts of the parameters):
\begin{gather}
  \Theta_{\lambda \mu} (x|Q) = \left[ \prod_{k \geq 0} G_{\lambda \mu}
    \left( x Q^k \right) G_{\mu \lambda} \left( \frac{q}{t} x^{-1}
      Q^{k+1} \right) \right] \exp \left[ \sum_{n \geq 1} \frac{x^n +
      \left( \frac{q}{t} Q
      \right)^n x^{-n}}{n(1-q^n)(1-t^{-n})(1-Q^n)} \right],\notag\\
  \Phi_{\lambda \mu} (x|Q) = \left[ \prod_{k \geq 0} G_{\lambda \mu}
    \left( \frac{q}{t} x Q^k\right) G_{\mu \lambda} \left( \frac{q}{t}
      x^{-1} Q^{k+1} \right) \right] \exp \left[ \sum_{n \geq 1}
    \left( \frac{q}{t} \right)^n \frac{x^n + Q^n
      x^{-n}}{n(1-q^n)(1-t^{-n})(1-Q^n)}
  \right],\label{eq:36}\\
  \overline{\Phi}_{\lambda \mu} (x|Q) = \left[ \prod_{k \geq 0}
    G_{\lambda \mu} \left( x Q^k\right) G_{\mu \lambda} \left( x^{-1}
      Q^{k+1} \right) \right] \exp \left[ \sum_{n \geq 1} \frac{x^n +
      Q^n x^{-n}}{n(1-q^n)(1-t^{-n})(1-Q^n)} \right].\notag
\end{gather}
These functions obey the following basic difference equations:
 \begin{align}
   \Theta_{\lambda \mu} \left(\frac{q}{t} x\Big| Q\right) &=
   \widetilde{\mathfrak{R}}^{(1)}_{\mu \lambda} (x^{-1}|Q)  \Theta_{\lambda \mu} (x|Q), \\
   \Phi_{\lambda \mu} \left( \frac{q}{t} x\Big| Q \right) &=
   \widetilde{\mathfrak{R}}^{(2)}_{\mu \lambda}
   \left(\frac{t}{q}x^{-1}\Big| Q\right) \Phi_{\lambda \mu} ( x|Q), \\
   \overline{\Phi}_{\lambda \mu} \left( \frac{q}{t} x\Big| Q\right) &=
   \widetilde{\mathfrak{R}}^{(3)}_{\mu \lambda} ( x^{-1}|Q)
   \overline{\Phi}_{\lambda \mu} ( x|Q).
 \end{align}

An explicit solution of the elliptic KZ equation is up to a periodic
function given by
\begin{framed}
  \begin{equation}
  \label{eq:43}
  \mathfrak{G}_{\mu_1 \cdots \mu_n }^{\lambda_1
    \cdots \lambda_n} (u_n|Q|
  \begin{smallmatrix}
    z_1, \ldots , z_n\\
    w_1, \ldots , w_n
  \end{smallmatrix}
  ) = \frac{\prod_{i,j}^n \Theta_{\lambda_i \mu_j} \left( \sqrt{\frac{q}{t}}
  \frac{z_i}{w_j} \Big| Q \right)}{\prod_{i<j} \Phi_{\lambda_j
  \lambda_i} \left( \frac{z_j}{z_i} \Big| Q \right) \prod_{k<l} \overline{\Phi}_{\mu_l
  \mu_k} \left(  \frac{w_l}{w_k} \Big| Q \right)}.
\end{equation}
\end{framed}
The three different propagators appear in the elliptic case, because
the pairings $\Psi\Psi$, $\Psi\Psi^{*}$ and $\Psi^{*}\Psi^{*}$ are all
different and cannot be expressed through a single function.

\subsection{Braiding, dual KZE and modular properties}
\label{sec:modular-properties}
The braiding operators $\mathcal{B}_{i,i+1}$, $\mathcal{B}^{j,j+1}$
defined by Eq.~\eqref{eq:33} are local in the sense that their
action depends only on the parameters of the two legs being
exchanged. Therefore, they are not modified by taking trace.

There is, however, one more braiding operator arising in the
elliptic case:
\begin{gather}
  \label{eq:48}
  \hat{\mathcal{B}}^{n,1} \mathfrak{G}_{\mu_1 \cdots \mu_n }^{\lambda_1
    \cdots \lambda_n} (u_n|Q|
  \begin{smallmatrix}
    z_1, \ldots , z_n\\
    w_1, \ldots , w_n
  \end{smallmatrix}
  ) \stackrel{\mathrm{def}}{=} \mathfrak{G}_{\mu_1 \cdots \mu_n }^{\lambda_1
    \cdots \lambda_n} (u_n|Q|
  \begin{smallmatrix}
    z_2, \ldots,  z_{n-1}, Q z_1, z_n\\
    w_1, \ldots , w_n
  \end{smallmatrix}
  ), \\
\hat{\mathcal{B}}_{n,1} \mathfrak{G}_{\mu_1 \cdots \mu_n }^{\lambda_1
    \cdots \lambda_n} (u_n|Q|
  \begin{smallmatrix}
    z_1, \ldots , z_n\\
    w_1, \ldots , w_n
  \end{smallmatrix}
  ) \stackrel{\mathrm{def}}{=} \mathfrak{G}_{\mu_1 \cdots \mu_n }^{\lambda_1
    \cdots \lambda_n} (u_n|Q|
  \begin{smallmatrix}
    z_1, \ldots ,  z_n\\
    w_2, \ldots  , w_{n-1}, Qw_1, w_n
  \end{smallmatrix}
  ).
\end{gather}
It acts in the following way on the solutions of the elliptic KZ
equation:
\begin{gather}
  \label{eq:49}
  \frac{\hat{\mathcal{B}}^{n,1} \mathfrak{G}_{\mu_1 \cdots \mu_n
  }^{\lambda_1 \cdots \lambda_n} (u_n|Q|
  \begin{smallmatrix}
    z_1, \ldots , z_n\\
    w_1, \ldots , w_n
  \end{smallmatrix}
  )}{\mathfrak{G}_{\mu_1 \cdots \mu_n
  }^{\lambda_1 \cdots \lambda_n} (u_n|Q|
  \begin{smallmatrix}
    z_1, \ldots , z_n\\
    w_1, \ldots , w_n
  \end{smallmatrix}
  )} = \mathcal{R}_{\lambda_n \lambda_1} \left( \frac{z_n}{z_1}
\right) \Upsilon_{q,t} \left( \frac{q}{t} \Big| \frac{z_1}{z_n}
\right) \prod_{i=1}^n \Upsilon_{q,t} \left( \sqrt{\frac{q}{t}} \Big| \frac{w_i}{z_1} \right) , \\
\frac{\hat{\mathcal{B}}_{n,1} \mathfrak{G}_{\mu_1 \cdots \mu_n }^{\lambda_1
  \cdots \lambda_n} (u_n|Q|
  \begin{smallmatrix}
    z_1, \ldots , z_n\\
    w_1, \ldots , w_n
  \end{smallmatrix}
  )}{\mathfrak{G}_{\mu_1 \cdots \mu_n }^{\lambda_1
  \cdots \lambda_n} (u_n|Q|
  \begin{smallmatrix}
    z_1, \ldots , z_n\\
    w_1, \ldots , w_n
  \end{smallmatrix}
  )} =  \frac{\Upsilon_{q,t} \left( 1 \Big|
    \frac{w_1}{w_n} \right)}{\mathcal{R}_{\mu_n \mu_1} \left(
    \frac{w_n}{w_1} \right)} \frac{1}{\prod_{j=1}^n \Upsilon_{q,t}
  \left( \sqrt{\frac{q}{t}}\Big| \frac{z_1}{w_j} \right)}.\label{eq:47}
\end{gather}
Notice that the anomaly factors in Eqs.~\eqref{eq:49},~\eqref{eq:47}
differ from those in Eqs.~\eqref{eq:34}\eqref{eq:6}. The complete set
of braiding transformations satisfies the \emph{affine} braid group
relations of type $\widehat{A}_n$. There is a set of distinguished
\emph{period} elements $\hat{\mathcal{P}}^k$, of the affine braid
group, which carry the variable $z_k$ over the full cycle under
the trace:
\begin{gather}
  \label{eq:50}
  \hat{\mathcal{P}}^k =
    \hat{\mathcal{B}}^{(k,k+1)}\hat{\mathcal{B}}^{(k+1,k+2)}\cdots
    \hat{\mathcal{B}}^{(n,1)}\hat{\mathcal{B}}^{(1,2)} \cdots
    \hat{\mathcal{B}}^{(k-1,k)}\\
     \hat{\mathcal{P}}_k =
    \hat{\mathcal{B}}_{(k,k+1)}\hat{\mathcal{B}}_{(k+1,k+2)}\cdots
    \hat{\mathcal{B}}_{(n,1)}\hat{\mathcal{B}}_{(1,2)} \cdots
    \hat{\mathcal{B}}_{(k-1,k)} \label{eq:51}
\end{gather}
The operator $\hat{\mathcal{P}}^k$ effectively shifts the variable
  $z_k$ by $Q$:
\begin{gather}
  \label{eq:52}
  \hat{\mathcal{P}}^k \mathfrak{G}_{\mu_1 \cdots \mu_n }^{\lambda_1
  \cdots \lambda_n} (u_n|Q|
  \begin{smallmatrix}
    z_1, \ldots , z_n\\
    w_1, \ldots , w_n
  \end{smallmatrix}
  ) = \mathfrak{G}_{\mu_1 \cdots \mu_n }^{\lambda_1
    \cdots \lambda_n} (u_n|Q|
  \begin{smallmatrix}
    z_1, \ldots, Q z_k,\ldots , z_n\\
    w_1, \ldots , w_n
  \end{smallmatrix}
  ),\\
  \hat{\mathcal{P}}_k \mathfrak{G}_{\mu_1 \cdots \mu_n }^{\lambda_1
  \cdots \lambda_n} (u_n|Q|
  \begin{smallmatrix}
    z_1, \ldots , z_n\\
    w_1, \ldots , w_n
  \end{smallmatrix}
  ) = \mathfrak{G}_{\mu_1 \cdots \mu_n }^{\lambda_1
    \cdots \lambda_n} (u_n|Q|
  \begin{smallmatrix}
    z_1, \ldots , z_n\\
    w_1, \ldots, Q w_k,\ldots , w_n
  \end{smallmatrix}
  ).\label{eq:53}
\end{gather}
One can evaluate the action of the period elements explicitly using
Eqs.~\eqref{eq:34},~\eqref{eq:6},~\eqref{eq:49} and~\eqref{eq:47} and
arrive at
\begin{framed}
  \begin{gather}
  \label{eq:54}
  \hat{\mathcal{P}}^k \mathfrak{G}_{\mu_1 \cdots \mu_n }^{\lambda_1
  \cdots \lambda_n} (u_n|Q|
  \begin{smallmatrix}
    z_1, \ldots , z_n\\
    w_1, \ldots , w_n
  \end{smallmatrix}
  ) = \frac{1}{\prod_{i=1}^n \Upsilon_{q,t}\left(
      \sqrt{\frac{q}{t}}\Big| \frac{z_k}{w_i} \right)}
  \frac{\prod_{i<k}\mathcal{R}^{\vee}_{\lambda_i \lambda_k}\left(
    \frac{z_i}{z_k} \right)}{\prod_{j>k}\mathcal{R}^{\vee}_{\lambda_k \lambda_j}\left( \frac{Q
        z_k}{z_j} \right)} \mathfrak{G}_{\mu_1 \cdots \mu_n
  }^{\lambda_1 \cdots \lambda_n} (u_n|Q|
  \begin{smallmatrix}
    z_1, \ldots , z_n\\
    w_1, \ldots , w_n
  \end{smallmatrix}
  ),\\
  \hat{\mathcal{P}}_k \mathfrak{G}_{\mu_1 \cdots \mu_n }^{\lambda_1
  \cdots \lambda_n} (u_n|Q|
  \begin{smallmatrix}
    z_1, \ldots , z_n\\
    w_1, \ldots , w_n
  \end{smallmatrix}
  ) = \frac{1}{\prod_{j=1}^n \Upsilon_{q,t}\left(
      \sqrt{\frac{q}{t}}\Big| \frac{z_j}{Q w_k} \right)}
  \frac{\prod_{j>k}
  \overline{\mathcal{R}}^{\vee}_{\mu_k \mu_j}\left( \frac{Q
        w_k}{w_j} \right)}{\prod_{i<k}\overline{\mathcal{R}}^{\vee}_{\mu_i \mu_k}\left(
    \frac{w_i}{w_k} \right)}  \mathfrak{G}_{\mu_1 \cdots \mu_n
  }^{\lambda_1 \cdots \lambda_n} (u_n|Q|
  \begin{smallmatrix}
    z_1, \ldots , z_n\\
    w_1, \ldots , w_n
  \end{smallmatrix}
  ),
\end{gather}
\end{framed}
\noindent
Here we introduced the \emph{dual} $\mathcal{R}$-matrices
differing from the DIM $\mathcal{R}$-matrix~\eqref{eq:9} by simple
scalar factors
\begin{gather}
  \label{eq:55}
  \mathcal{R}^{\vee}_{\lambda \mu}(x) =
  \frac{f_{\lambda}}{f_{\mu}}\mathcal{R}_{\lambda \mu}\left( x \right)
  \Upsilon_{q,t}\left( \frac{q}{t} \Big| x^{-1} \right),\qquad \qquad
  \mathcal{R}^{\vee}_{\mu\lambda }(x^{-1}) =
  (\mathcal{R}^{\vee}_{\lambda \mu }(x))^{-1},\\
  \overline{\mathcal{R}}^{\vee}_{\lambda \mu}(x) = \left( \frac{q}{t}
  \right)^{|\mu|-|\lambda|}
  \frac{f_{\mu}}{f_{\lambda}}\mathcal{R}_{\lambda \mu}\left( x \right)
  \Upsilon_{q,t}\left( 1 \Big| x \right),\qquad \qquad
  \overline{\mathcal{R}}^{\vee}_{\mu\lambda }(x^{-1}) =
  (\overline{\mathcal{R}}^{\vee}_{\lambda \mu }(x))^{-1}
\end{gather}
Eqs.~\eqref{eq:54} are \emph{dual} KZE satisfied by the trace of
intertwiners. They are very similar to the $(q,t)$-KZ
equations~\eqref{eq:30},~\eqref{eq:31}, except for the shifts in the
dual picture which are not $\frac{q}{t}$, but $Q$! Moreover, the
$\mathcal{R}$-matrices in the dual equation are not elliptic, but
trigonometric. It is natural to compare this with the analogous
results for quantum affine
algebras~\cite{Etin, Sun} where both the
$q$-KZ and dual $q$-KZ equations were elliptic.

Let us also briefly discuss the modular properties of solutions that we
have just obtained. We focus on the simplest case of $n=1$, i.e.\
$\mathfrak{G}_{\mu
  }^{\lambda} (u|Q|
  \begin{smallmatrix}
    z\\
    w
  \end{smallmatrix}
  )$, and then give some comments about the general solutions.  We
  use the following identity\footnote{The infinite product in the
    r.h.s.\ requires some regularization.}:
\begin{equation}
  \label{eq:67}
  \Theta_{\lambda \mu}(x|Q) = \prod_{k \geq 1} (1 -
  Q^k)^{-|\lambda|-|\mu|} \prod_{(i,j)\in \lambda} \theta_Q (x
  q^{\lambda_i - j} t^{i - \mu_j^{\mathrm{T}}+1}) \prod_{(i,j)\in \mu}
  \theta_Q (x q^{j -\mu_i - 1} t^{i - \lambda_j^{\mathrm{T}}})
  \prod_{i,j \geq 0} \frac{\theta_Q(x q^i t^{-j})}{\prod_{k\geq 1} (1
    - Q^k)}
\end{equation}
where $\theta_Q(x) = \prod_{k \geq 0} (1 - Q^{k+1})(1 - Q^k x)(1 -
x^{-1} Q^{k+1})$. We use the modular properties of the theta-function and the Dedekind function:
\begin{gather}
  \label{eq:68}
  \theta_{\tilde{Q}}(\tilde{x}) = (-i \tau)^{\frac{1}{2}} e^{\frac{i
      \pi}{\tau} \left( a + \frac{1}{2} - \frac{\tau}{2} \right)^2}
  \theta_Q(x)\\
  \prod_{k \geq 1} \left( 1 - \tilde{Q}^k \right) = (-i \tau)^{\frac{1}{2}} Q^{\frac{1}{24}} \tilde{Q}^{-\frac{1}{24}} \prod_{k \geq 1} \left( 1 - Q^k \right),
\end{gather}
where
\begin{gather}
  \label{eq:69}
  Q = e^{2\pi i \tau}, \qquad \tilde{Q} = e^{- \frac{2 \pi i}{\tau}}\\
  x = e^{2\pi i a}, \qquad \tilde{x} = e^{2\pi i \frac{a}{\tau}}
\end{gather}
We note that $\Theta_{\varnothing \varnothing}(x \vert Q)$ is nothing
but the inverse of the double elliptic gamma function $G_2(x \vert q,
t^{-1}, Q)$ defined by
\begin{equation}
  \label{eq:76}
  G_2(x \vert q_0, q_1, q_2) \stackrel{\mathrm{def}}{=} \prod_{i,j,k=0}^\infty ( 1 - x q_0^i q_1^j q_2^k)
(1 - x^{-1} q_0^{i+1} q_1^{j+1} q_2^{k+1}).
\end{equation}
This is the third member of a hierarchy of meromorphic
functions\footnote{These functions
  $$
  G_r(x|\vec{q}) = (x|\vec{q})_{\infty}^{(-1)^r} (x^{-1} q_0\cdots
  q_r|\vec{q})_{\infty}
  $$
  are defined using the multiple
  $q$-Pochhammer symbol
  $$
  (x| \vec{q})_{\infty} = \prod_{j_0, \ldots, j_r} \left( 1 - x
    q_0^{j_0}\cdots q_r^{j_r} \right).
  $$
} $G_r(x \vert q_0, \cdots, q_{r})$ starting from a Jacobi theta
function $\vartheta_0(z, \tau) = G_0(e^{2\pi i z} \vert e^{2\pi i
  \tau})$.  The elliptic gamma function $\Gamma(z, \tau, \sigma) =
G_1(e^{2\pi i z} \vert e^{2\pi i \tau} , e^{2\pi i\sigma})$ was
introduced in \cite{Ruijs} and implicitly appeared in the formula of
the free energy of the eight vertex model \cite{Baxter}.  It also
appears in the solutions to the elliptic $q$-KZB equations \cite{FTV,
  FV1}.  The modular property of the elliptic gamma function is
discussed in \cite{FV2}.  The function $G_r(x \vert q_0, \cdots,
q_{r})$ are called multiple elliptic gamma functions and regarded as
an elliptic generalization of the Barnes multiple gamma function
\cite{Nishi}.  The useful formulas for $G_r(x \vert q_0, \cdots,
q_{r})$ including the modular property are summarized in
\cite{Naru}. More recently the full partition function of 5d
$\mathcal{N}=2^*$ gauge theory with a hypermultiplet in the adjoint
representation was expressed using the double elliptic gamma function
$G_2(x \vert q_0, q_1, q_2)$ \cite{Lockhart:2012vp, Qiu:2015rwp}.
Finally there is a closely related function called multiple sine
function $S_r(z \vert q_0, \cdots, q_{r})$.  It is amusing that some
solutions of the $q$-KZ equation for XXZ model (or
$U_q(\widehat{\mathfrak{sl}}_2)$) are given by the double sine
function \cite{Jimbo:1996ss}.

Solution~\eqref{eq:43} to the elliptic $(q,t)$-KZ equations have quite
involved modular behaviour because of the factor $\Theta_{\varnothing
  \varnothing}(x \vert Q)$. However, one can obtain a relatively
simple transformation law for the \emph{ratio} of two solutions:
\begin{multline}
  \label{eq:70}
 \frac{\mathfrak{G}_{\mu
  }^{\lambda} (u|\tilde{Q},\tilde{q},\tilde{t}|
  \begin{smallmatrix}
    \tilde{z}\\
    \tilde{w}
  \end{smallmatrix}
  )}{\mathfrak{G}_{\varnothing}^{\varnothing} (u|\tilde{Q},\tilde{q},\tilde{t}|
  \begin{smallmatrix}
    \tilde{z}\\
    \tilde{w}
  \end{smallmatrix}
  )} = \frac{\Theta_{\lambda
      \mu}\left(\sqrt{\frac{\tilde{q}}{\tilde{t}}} \frac{\tilde{z}}{\tilde{w}}
      |\tilde{Q},\tilde{q},\tilde{t}\right)}{\Theta_{\varnothing
      \varnothing}\left(\sqrt{\frac{\tilde{q}}{\tilde{t}}}
      \frac{\tilde{z}}{\tilde{w}}|\tilde{Q},\tilde{q},\tilde{t}\right)}
  = e^{-\frac{i\pi}{12} \left( \tau + \frac{1}{\tau} \right)
    (|\lambda|+ |\mu|)}\times\\
  \times \exp \left[ \frac{i \pi}{\tau} \sum_{(i,j)\in
      \lambda} \left( a - b  + \epsilon_2\left(j - \lambda_i + \frac{1}{2}\right) + \epsilon_1
      \left(\mu_j^{\mathrm{T}}-i+\frac{1}{2}\right) + \frac{1}{2} - \frac{\tau}{2}
    \right)^2 \right]\times\\
  \times   \exp \left[ \frac{i \pi}{\tau} \sum_{(i,j)\in
      \mu} \left( a - b+ \epsilon_2 \left(\mu_i-j + \frac{1}{2}\right) + \epsilon_1
      \left(i-\lambda_j^{\mathrm{T}} - \frac{1}{2}\right) + \frac{1}{2} - \frac{\tau}{2}
    \right)^2 \right] \frac{\Theta_{\lambda
      \mu}\left(\sqrt{\frac{q}{t}} \frac{z}{w}
      |Q,q,t\right)}{\Theta_{\varnothing
      \varnothing}\left(\sqrt{\frac{q}{t}}
      \frac{z}{w}|Q,q,t\right)}
\end{multline}
where
\begin{gather}
  \label{eq:71}
  q = e^{-2\pi i \epsilon_2}, \qquad \tilde{q} = e^{-2\pi i
    \frac{\epsilon_2}{\tau}}\\
  t = e^{2\pi i \epsilon_1}, \qquad \tilde{t} = e^{2\pi i
    \frac{\epsilon_1}{\tau}}\\
  z= e^{2\pi i a}, \qquad \tilde{z} = e^{2\pi i \frac{a}{\tau}}\\
  w= e^{2\pi i b}, \qquad \tilde{w} = e^{2\pi i \frac{b}{\tau}}
\end{gather}
We can see that the modular transformation is diagonal and given by
multiplication with a simple factor. As usual, one can restore the
exact modular invariance by adding an antiholomorphic term in the
solution. The solution will then satisfy the holomorphic anomaly
equation.

For higher-point correlators, i.e.\ for $n \geq 1$ the transformation
properties of the solution are more involved. In particular, the
$\Phi$ and $\bar{\Phi}$ propagators are not modular invariant. The
transformation matrix is nontrivial and is no longer given by a simple
overall factor. It would be very interesting to compute this matrix
exactly.

Let us mention that our results are parallel to the properties of the
M-strings partition function~\cite{M-strings}. In fact, solutions
of the elliptic $(q,t)$-KZ equations provide the basic building block of the
partitions functions considered in~\cite{M-strings}. We hope that the
$(q,t)$-KZE can be also naturally understood within this context.

We would like to point out an interesting observation given
in~\cite{FV1,FV2} that the group of modular transformations
acting on solutions to the elliptic $q$-KZ is generalized from $SL(2,
\mathbb{Z})$ to $SL(3,\mathbb{Z})$. It is tempting to assume that in
the DIM case this symmetry is enhanced even further. This can be seen
from the fact that the modular group acting on the function $G_r$ is
in general $SL(r+2, \mathbb{Z})$. Since the solution~(\ref{eq:43}) is
related to $G_2$ (see Eq.~(\ref{eq:76})) it naturally transforms under
$SL(4, \mathbb{Z})$.

\section{Conclusion}

We have considered a natural question within the context of the
network models with underlying DIM symmetry: the way this symmetry is
actually realized on the correlators.  Since the original studies of
WZW model, it is traditional to describe this action in terms of the
Knizhnik-Zamolodchikov and Knizhink-Zamolodchikov-Bernard equations,
which express the variation of conformal blocks under the change of
moduli through the action of the symmetry.  Equations of similar type
for the network models were already considered in \cite{Okounkov-Smirnov}, but
the emphasis there was on the dependence on K\"ahler structures, while
the ordinary KZ equation was largely ignored.  The reason for this is
probably that the models currently studied have the
$\widehat{\widehat{\mathfrak{u}}}_1$ symmetry, which is essentially
Abelian, and this leads to an extraordinary simplification of the
ordinary KZE making it look too trivial.  Still, the equation exists,
should be written and investigated, and this is the purpose of the
present text.

From two most popular approaches to the KZE, the free field and the
abstract group theory approaches, we took the former one.  The latter
approach is also interesting, but considerably different and will be
addressed elsewhere.  In the free field formalism \cite{GMMOS}, the
main origin of non-trivial (transcendental) solutions to the KZE are
insertions of peculiar screening charges, which commute with the
chiral algebra.  While screening charges exist in all network models
\cite{NagoyaITEP}, in the case of
$\widehat{\widehat{\mathfrak{u}}}_1$ they commute only with the
Virasoro/$W_N$-operators but not with the generators of the DIM
algebra.  In terms of the balanced network models, this means that the
KZE appears only for the single horizontal line networks, where
screening charges are not allowed since they are associated with the
vertical edges connecting different horizontal lines.  The
corresponding averages (network conformal blocks) are algebraic,
direct counterparts of $\prod_{i<j}(z_i-z_j)^{\alpha_i\alpha_j}$
rather than of the Schechtman-Varchenko ``hypergeometric'' functions,
obtained by integration over some of the $z$-variables.  This is the
property shared by all the solutions to the ordinary KZE for the WZW
models with $\widehat{\widehat{\mathfrak{u}}}_1$ symmetry.

Let us notice that the combinations of intertwiners which we consider
in this paper also appear as building blocks of the Nekrasov partition
functions. In particular, the trigonometric block corresponds to the
bifundamental matter contribution in $5d$ theory (see
Fig.~\ref{Nekr}), while the elliptic block gives rise to the $6d$
version. By the spectral duality, the $6d$ partition function with
bifundamental matter can be thought of as the $5d$ adjoint one. The
crucial point is that though the individual conformal blocks satisfy
the $(q,t)$-KZ equations, their convolution featuring in the Nekrasov
function does not. The reason is that the $\mathcal{R}$-matrices in
the r.h.s.\ of the $(q,t)$-KZ equations for the two horizontal strands
in Fig.~\ref{Nekr} do not cancel.

Still, even in the simplest Abelian case there are many questions to
ask and directions to explore.

For example, in the elliptic generalization, where the modular group
acts, we provide only preliminary results about the emerging modular
matrices. These matrices might be useful in the construction of
refined knot invariants. Another kind of questions is about emerging
integrable strictures.

As is well-known, the solutions to KZE in the semiclassical limit are
related to solutions of the Bethe equations for the corresponding
quantum integrable systems~\cite{Tarasov-Varchenko}.  The same
relation should be valid in the DIM case. It will be studied in
future publications.

However, the most interesting directions are the study of generic
representations (associated with plane rather than ordinary
partitions) and most importantly the lifting from
$\widehat{\widehat{\mathfrak{u}}}_1$ to
$\widehat{\widehat{\mathfrak{g}}}$ with a simple Lie algebra
$\mathfrak{g}$, where generalizations of the true non-Abelian KZE
emerge, even for the Fock representations.  In the free field
approach, this requires lifting of the entire technique of
\cite{GMMOS} to the level of DIM, which is straightforward but
cumbersome. Further lifting to generic (plane-partition)
representations is a real challenge requiring the ``double loop''
chiral free fields, which is a separate very interesting problem with
many promising ideas coming from attempts in different branches of
theoretical physics. This paper is just the first step on the long
road to full understanding of symmetries of the network models and
their realization through the variety of KZ and KZB equations.

\section*{Acknowledgements}

A.M.'s and Y.Z. acknowledge the hospitality of Nagoya University
in the course of Nagoya-ITEP collaboration.
We very much appreciate illuminating discussions with S.Nawata
at the beginning of this project. Y.Z. would like to thank I. Frenkel for discussions.

\bigskip

Our work is supported in part by Grants-in-Aid for Scientific Research
(\# 24540210) (H.A.), (\# 15H05738) (H.K.), for JSPS Fellow (\#
26-10187) (Y.O.), (\# S16124) (Al.Mor.)  and JSPS Bilateral Joint Projects (JSPS-RFBR
collaboration) ``Exploration of Quantum Geometry via Symmetry and
Duality'' from MEXT, Japan. It is also partly supported by grants
15-31-20832-Mol-a-ved (Al.Mor.), 15-31-20484-Mol-a-ved (Y.Z.),
16-32-60047-Mol-a-dk (And.Mor), by RFBR grants 16-01-00291 (A.Mir.),
15-01-09242 (An.Mor.) and 14-01-00547 (Y.Z.), by joint grants
17-51-50051-YaF, 15-51-52031-NSC-a, 16-51-53034-GFEN,
16-51-45029-IND-a. The work of Y.Z. was supported in part by INFN and
by the ERC Starting Grant 637844-HBQFTNCER.


\begin{thebibliography}{12}

\bibitem{AGT} L.~Alday, D.~Gaiotto and Y.~Tachikawa,
  Lett.\ Math.\ Phys.\ {\bf 91} (2010) 167-197, arXiv:0906.3219\\
  N.~Wyllard,
  JHEP {\bf 0911} (2009) 002, arXiv:0907.2189\\
  A.~Mironov and A.~Morozov, Nucl.\ Phys.\ {\bf B825} (2009) 1-37,
  arXiv:0908.2569

\bibitem{AGT5d} H.~Awata and H.~Kanno,
  JHEP {\bf 0907} (2009) 076
arXiv:0905.0184\\
H.~Awata and Y.~Yamada,
  JHEP {\bf 1001} (2010) 125,
arXiv:0910.4431;
  Prog.\ Theor.\ Phys.\  {\bf 124} (2010) 227,
arXiv:1004.5122\\
S. Yanagida, arXiv:1005.0216\\
H.~Awata, H.~Fuji, H.~Kanno, M.~Manabe and Y.~Yamada,
  Adv.\ Theor.\ Math.\ Phys.\  {\bf 16} (2012) no.3,  725
arXiv:1008.0574\\
H.~Kanno and Y.~Tachikawa,
  JHEP {\bf 1106} (2011) 119
arXiv:1105.0357\\
A. Mironov, A. Morozov, S. Shakirov and A. Smirnov, Nucl. Phys. {\bf B855} (2012) 128, arXiv:1105.0948\\
H.~Kanno and M.~Taki,
  JHEP {\bf 1205} (2012) 052
  arXiv:1203.1427\\
F. Nieri, S. Pasquetti and F. Passerini, arXiv:1303.2626\\
  F. Nieri, S. Pasquetti, F. Passerini and A. Torrielli, arXiv:1312.1294\\
M.-C. Tan, JHEP {\bf 12} (2013) 031, arXiv:1309.4775; arXiv:1607.08330\\
H. Itoyama, T.Oota and R. Yoshioka, arXiv:1408.4216, arXiv:1602.01209\\
  A. Nedelin and M. Zabzine, arXiv:1511.03471\\
  R. Yoshioka, arXiv:1512.01084\\
    Y.~Ohkubo, H.~Awata and H.~Fujino,
  arXiv:1512.08016\\
S. Pasquetti, arXiv:1608.02968

\bibitem{CFT} A. Belavin, A. Polyakov and A. Zamolodchikov, Nucl. Phys. {\bf B241} (1984) 333-380\\
A. Zamolodchikov, Al. Zamolodchikov, {\sl Conformal field theory and critical phenomena in 2d systems}, 2009\\
L. Alvarez-Gaume, Helvetica Physica Acta {\bf 64} (1991) 359-526\\
P. Di Francesco, P. Mathieu and D. Senechal, {\sl Conformal Field Theory}, Springer, 1996

\bibitem{ADHM} M. Atiyah, V. Drinfeld and N. Hitchin, Yu. Manin, Phys. Lett. {\bf A65} (1978) 185

\bibitem{LMNS} G. Moore, N. Nekrasov and S. Shatashvili, Commun. Math. Phys. {\bf 209} (2000) 97-121, hep-th/9712241;
{\it ibid.} 77-95, hep-th/9803265\\
A. Losev, N. Nekrasov and S. Shatashvili, Nucl. Phys. {\bf B534} (1998) 549-611, hep-th/9711108; hep-th/9801061

\bibitem{Nakajima}  H. Nakajima, 
Duke Math. J. {\bf 76} (1994), 365-416; 
{\it ibid.} {\bf 91} (1998), 515-560;
{\sl Lectures on Hilbert schemes of points on surfaces}, University Lecture Series, 18. American Mathematical
Society, Providence, RI, 1999

\bibitem{Nekrasov} N. Nekrasov, Adv. Theor. Math. Phys. {\bf 7} (2004) 831-864, hep-th/0206161

\bibitem{GW} M. Gromov, 
Inv. Math. {\bf 82} (1985) 307-347\\
M. Dine, N. Seiberg, X. Wen and E. Witten, 
Nucl.Phys. {\bf B278} (1986) 769\\
E. Witten, 
Comm. Math. Phys. {\bf 118} (1988) 411-449

\bibitem{NO} N.Nekrasov and A.Okounkov, hep-th/0306238

\bibitem{Aganagic-triality} M.~Aganagic, N.~Haouzi, C.~Kozcaz and
  S.~Shakirov, 
  arXiv:1309.1687 \\
M.~Aganagic, N.~Haouzi and S.~Shakirov, 
  arXiv:1403.3657\\
M.~Aganagic and N.~Haouzi, arXiv:1506.04183

\bibitem{spec-dual} E. Mukhin, V. Tarasov and A. Varchenko, 
math/0510364;
Adv. Math. {\bf 218} (2008) 216-265, math/0605172\\
A. Mironov, A. Morozov, Y. Zenkevich and A. Zotov, JETP Lett. {\bf 97} (2013) 45, arXiv:1204.0913\\
A. Mironov, A. Morozov, B. Runov, Y. Zenkevich and A. Zotov, Lett. Math. Phys. {\bf 103} (2013) 299,
arXiv:1206.6349; JHEP {\bf 1312} (2013) 034, arXiv:1307.1502\\
L. Bao, E. Pomoni, M. Taki and F. Yagi, JHEP {\bf 1204} (2012) 105, arXiv:1112.5228

\bibitem{Okounkov-Smirnov} A. Okounkov and A. Smirnov, 1602.09007

\bibitem{DIM} J. Ding, K. Iohara, 
Lett. Math. Phys. {\bf 41} (1997) 181-193, q-alg/9608002\\
K. Miki, J. Math. Phys. {\bf 48} (2007) 123520\\
B. Feigin and A. Tsymbaliuk, Kyoto J. Math. {\bf 51} (2011) 831-854, arXiv:0904.1679\\
B. Feigin, E. Feigin, M. Jimbo, T. Miwa and E. Mukhin, Kyoto J. Math. {\bf 51} (2011) 337-364, arXiv:1002.3100\\
B.~Feigin, K.~Hashizume, A.~Hoshino, J.~Shiraishi and S.~Yanagida, J.~Math.~Phys. \textbf{50} (2009) 095215, arXiv:0904.2291\\
B. Feigin, A. Hoshino, J. Shibahara, J. Shiraishi and S. Yanagida, arXiv:1002.2485\\
B. Feigin, E. Feigin, M. Jimbo, T. Miwa and E. Mukhin, Kyoto J. Math. {\bf 51} (2011) 365-392, arXiv:1002.3113\\
  H.~Awata, B.~Feigin, A.~Hoshino, M.~Kanai, J.~Shiraishi and S.~Yanagida,
  RIMS k\={o}ky\={u}roku {\bf 1765} (2011) 12-32;
  arXiv:1106.4088\\
B. Feigin, M. Jimbo, T. Miwa and E. Mukhin, Kyoto J. Math. {\bf 52} (2012) 621-659, arXiv:1110.5310; arXiv:1502.07194

\bibitem{MMZpagoda} A.~Mironov, A.~Morozov, Y.~Zenkevich, Phys. Lett. {\bf B762} (2016) 196-208, arXiv:1603.05467

\bibitem{NagoyaITEP} H. Awata, H. Kanno, T. Matsumoto, A. Mironov, A. Morozov, An. Morozov, Y. Ohkubo and Y. Zenkevich, JHEP {\bf 07} (2016) 103, arXiv:1604.08366

\bibitem{AFS} H. Awata, B. Feigin and J. Shiraishi, arXiv:1112.6074

\bibitem{DF} B.Feigin and D.Fuks, Funct. Anal. Appl. {\bf 16} (1982)
114-126 (Funkt. Anal. Pril. {\bf 16} (1982) 47-63)\\
Vl. Dotsenko and V. Fateev, Nucl. Phys. {\bf B240} (1984) 312-348

\bibitem{MMSh} A. Mironov, A. Morozov and Sh. Shakirov,
JHEP {\bf 02} (2010) 030, arXiv:0911.5721;
Int. J. Mod. Phys. {\bf A25} (2010) 3173-3207, arXiv:1001.0563;
JHEP {\bf 1103} (2011) 102, arXiv:1011.3481

\bibitem{bn} A. Iqbal, C. Kozcaz and C. Vafa, JHEP {\bf 0910} (2009) 069, hep-th/0701156\\
H.~Awata and H.~Kanno,
JHEP {\bf 0505} 039 (2005),
  hep-th/0502061;
  Int.\ J.\ Mod.\ Phys.\ A {\bf 24} (2009) 2253, arXiv:0805.0191
H.~Awata and H.~Kanno,
  J.\ Geom.\ Phys.\  {\bf 64} (2013) 91
arXiv:0903.5383

\bibitem{WZNW}  J. Wess and B. Zumino, Phys. Lett. {\bf B37} (1971) 95\\
S. Novikov, Usp. Mat. Nauk {\bf 37} (1982) 3\\
E. Witten, 
Comm. Math. Phys. {\bf 92} (1984) 455\\
A. Polyakov and P. Wiegmann, 
Phys. Lett. {\bf B131} (1983) 121; 
Phys. Lett. {\bf B141} (1984) 223

\bibitem{GMMOS} M. Wakimoto, 
Commun. Math. Phys. {\bf 104} (1986) 605-609\\
A. Gerasimov, A. Marshakov, A. Morozov, M. Olshanetsky, S. Shatashvili,
Int. J. Mod. Phys. A5 (1990) 2495-2589\\
V. Dotsenko, 
Nucl. Phys. {\bf B338} 747 (1990); 
Nucl. Phys. {\bf B358} (1991) 547\\
B. Feigin and E. Frenkel 
Phys. Lett. {\bf B246} (1990) 75-81

\bibitem{KZ} V. Knizhnik and A. Zamolodchikov, 
Nucl. Phys. {\bf B247} (1984) 83

\bibitem{FR} I. Frenkel and N. Reshetikhin, 
Comm. Math. Phys. {\bf 146} (1992) 1-60

\bibitem{KZB} D. Bernard, 
Nucl. Phys. {\bf B303} (1988) 77-93; {\it ibid.}
{\bf B309} (1988) 145-174\\
G. Felder and A. Varchenko, Int. Math. Res. Notices (1995) 221-233, hep-th/9502165

\bibitem{Etin} P. Etingof and A. Varchenko, math/9907181; math/0302071\\
P. Etingof, O. Schiffmann and A. Varchenko, 
arXiv:math/0207157

\bibitem{RF}  N. Reshetikhin and L.D. Faddeev, 
Theor. Math. Phys. {\bf 56} (1983) 323-343

\bibitem{Sun} Yi Sun, 
arXiv:1609.09038

\bibitem{Sm} A. Smirnov, arXiv:1302.0799; arXiv:1404.5304

\bibitem{NagoyaITEP3} H.~Awata, H.~Kanno, A.~Mironov, A.~Morozov, A.~Morozov, Y.~Ohkubo and Y.~Zenkevich, arXiv:1611.07304

\bibitem{O} A. Okounkov, 1512.07363

\bibitem{S} A. Smirnov, 1612.01048

\bibitem{MMZ} A.~Morozov and Y.~Zenkevich,
JHEP {\bf 1602} (2016) 098,  arXiv:1510.01896\\
A.~Mironov, A.~Morozov and Y.~Zenkevich, Phys. Lett. {\bf B756} (2016) 208-211, arXiv:1512.06701; JHEP, {\bf 05} (2016) 121, arXiv:1603.00304; Phys. Lett. {\bf B762} (2016) 196-208, arXiv:1603.05467

\bibitem{NagoyaITEP2} H.~Awata, H.~Kanno, A.~Mironov, A.~Morozov, A.~Morozov, Y.~Ohkubo and Y.~Zenkevich,
JHEP {\bf 10} (2016) 047, arXiv:1608.05351

\bibitem{SchV} V. Schechtman and A. Varchenko, 
Lett. Math. Phys. {\bf 20} (1990) 279; 
Inv. Math.  {\bf 106} (1991) 139\\
H. Awata, A. Tsuchiya and Y. Yamada, 
Nucl. Phys. {\bf B365} (1991) 680\\
H. Awata, 
Prog. Theor. Phys. Suppl. {\bf 110} (1992) 303, hep-th/9202032

\bibitem{Zam} A.B. Zamolodchikov, Comm. Math. Phys. {\bf 69} (1979) 165

\bibitem{FJMM} B.~Feigin, M.~Jimbo, T.~Miwa and E.~Mukhin, arXiv:1603.02765

\bibitem{Nakajima:2003pg}
  H.~Nakajima and K.~Yoshioka,
  Inv.\ Math.\  {\bf 162} (2005) 313,
math/0306198

\bibitem{ellDIM} Y. Saito, 
    arXiv:1301.4912;
SIGMA {\bf 10} (2014) 021, arXiv:1305.7097;
arXiv:1309.7094

\bibitem{Ruijs}
S.N.M.~Ruijsenaars,
  J.\ Math.\ Phys {\bf 38} (1997) 1069

\bibitem{Baxter}
 R.J.~Baxter,
Ann.\ Phys.  {\bf 70} (1972) 193

\bibitem{FTV}
G.~Felder, V.~Tarasov and A.~Varchenko,
Amer.\ Math.\ Soc.\ Transl. {\bf 180} (1997) 45;
  Int.\ J.\ Math. {\bf 10} (1999) 943, q-alg/9705017

\bibitem{FV1}
G.~Felder and A.~Varchenko,
math.QA/9809139


\bibitem{FV2}
G.~Felder and A.~Varchenko,
  Adv.\ Math. {\bf 156} (2000) 44, math.QA/9907061

\bibitem{Nishi}
M.~Nishizawa,
   J.\ Phys.\ A {\bf 34} (2001) 7411

\bibitem{Naru}
A.~Narukawa,
  Adv.\ Math. {\bf 189} (2004) 247, math.QA/0306164

\bibitem{Lockhart:2012vp}
  G.~Lockhart and C.~Vafa,
  arXiv:1210.5909

\bibitem{Qiu:2015rwp}
  J.~Qiu, L.~Tizzano, J.~Winding and M.~Zabzine,
  JHEP {\bf 1603} (2016) 193,
arXiv:1511.06304\\
R. Lodin, F. Nieri and M. Zabzine, arXiv:1703.04614

\bibitem{Jimbo:1996ss}
  M.~Jimbo and T.~Miwa,
  J.\ Phys.\ A {\bf 29} (1996) 2923,
hep-th/9601135

\bibitem{M-strings} B.~Haghighat, A.~Iqbal, C.~Kozcaz, G.~Lockhart and C.~Vafa,
  Commun.\ Math.\ Phys.\  {\bf 334} (2015) 779,
arXiv:1305.6322\\
B.~Haghighat, C.~Kozcaz, G.~Lockhart and C.~Vafa,
  Phys.\ Rev.\ {\bf D89} (2014) 046003
arXiv:1310.1185

\bibitem{Tarasov-Varchenko} A. Beilinson and V. Drinfeld, {\sl Quantization of Hitchin's fibration and Langlands
program}, in: {\sl Algebraic and Geometric Methods in Mathematical Physics} (A.
Boutet de Monvel and V. A. Marchenko, editors), Mathematical Physics Studies,
vol. 19, 3-7. Kluwer Academic Publishers, Netherlands, 1996;
{\sl Quantization of Hitchin's integrable system
and Hecke eigensheaves}, http://www.math.uchicago.edu/~mitya/langlands/hitchin/BD-hitchin.pdf\\
B. Feigin, E. Frenkel and N. Reshetikhin, Commun. Math. Phys. {\bf 166} (1994) 27-62, hep-th/9402022

\end{thebibliography}
\end{document}